\documentclass[fleqn,usenatbib]{mnras}
\usepackage{float}
\usepackage{placeins}
\usepackage{newtxtext,newtxmath}

\usepackage{caption}
\usepackage{etoolbox}
\robustify\cite

\usepackage[switch]{lineno}

\usepackage[T1]{fontenc}
\usepackage{orcidlink}

\DeclareRobustCommand{\VAN}[3]{#2}
\let\VANthebibliography\thebibliography
\def\thebibliography{\DeclareRobustCommand{\VAN}[3]{##3}\VANthebibliography}

\newcommand{\tcoolsh}{t_{\rm cool}}
\newcommand{\tff}{t_{\rm ff}}

\usepackage{graphicx }	

\usepackage{amsmath, amssymb}	
\usepackage{tabularx}

\title[Three-phase evolution of the MW]{From protogalaxy through thick and thin: Why did the Milky Way evolve in three kinematic phases?}

\author[O. Myrtaj et al.]{%
Olti Myrtaj,$^{1}$\thanks{E-mail: \href{mailto:omyrtaj@uci.edu}{omyrtaj@uci.edu}}
James S. Bullock,$^{1,2}$
Michael Boylan-Kolchin,$^{3,4,5}$
Vedant Chandra\orcidlink{0000-0002-0572-8012},$^{6}$ \newauthor
Claude-Andr\'e Faucher-Gigu\`ere\orcidlink{0000-0002-4900-6628},$^{7}$
Robert Feldmann,$^{8}$
Pratik J. Gandhi\orcidlink{0000-0003-0965-605X},$^{9}$
Francisco J. Mercado,$^{10}$ \newauthor
Jorge Moreno\orcidlink{0000-0002-3430-3232},$^{10,11}$
Jonathan Stern,$^{12}$
Andrew Wetzel,$^{13}$
Luke Y. Xia\orcidlink{0009-0002-4568-2423}$^{1}$
\\
$^{1}$Department of Physics and Astronomy, University of California, Irvine, CA 92697, USA\\
$^{2}$Department of Physics and Astronomy, University of Southern California, Los Angeles, CA 90089, USA\\
$^{3}$Department of Astronomy, The University of Texas at Austin, Austin, TX 78712, USA\\
$^{4}$Cosmic Frontier Center, The University of Texas at Austin, Austin, TX 78712, USA\\
$^{5}$Weinberg Institute for Theoretical Physics, The University of Texas at Austin, Austin, TX 78712, USA\\
$^{6}$Center for Astrophysics | Harvard \& Smithsonian, 60 Garden St, Cambridge, MA 02138, USA\\
$^{7}$Department of Physics and Astronomy and CIERA, Northwestern University, 2145 Sheridan Road, Evanston, IL 60208, USA\\
$^{8}$Department of Astrophysics, University of Zurich, Winterthurerstrasse 190, 8057 Zurich, Switzerland\\
$^{9}$Department of Astronomy, Yale University, New Haven, CT 06520, USA\\
$^{10}$Department of Physics and Astronomy, Pomona College, Claremont, CA 91711, USA\\
$^{11}$Carnegie Observatories, 813 Santa Barbara St., Pasadena, CA 91101, USA\\
$^{12}$School of Physics and Astronomy, Tel Aviv University, Tel Aviv 69978, Israel\\
$^{13}$Department of Physics and Astronomy, University of California, Davis, CA 95616, USA
}

\pubyear{2026}

\begin{document}
\label{firstpage}
\pagerange{\pageref{firstpage}--\pageref{lastpage}}
\maketitle

\begin{abstract}
APOGEE and Gaia data suggest the Milky Way's kinematic structure evolved through three distinct phases: a disordered protogalaxy, which ``spun up'' into a thick stellar disk, and then ``cooled down'' to a final cold, thin stellar disk. We use a suite of FIRE-2 cosmological zoom-in simulations of Milky Way-mass galaxies to demonstrate that the same three phases arise in simulations, and we study their physical origin. In our simulations, the early protogalaxy phase occurs when the rate of cool gas ($T \leq 10^4$ K) converting into stars is low, star formation is bursty, and the baryonic mass ``sloshes'' within the host potential with respect to the center of mass. The gas begins to spin coherently after sloshing ends, followed by the spin-up of young stars. The central potential is least concentrated just prior to gas spin-up. This second, thick disk phase coincides with the highest rate of cool gas converting into stars, though star formation remains bursty. The final transition to the thin disk phase occurs when the inner circumgalactic medium virializes, and is associated with steady star formation and intermediate consumption rates of cool gas converting into stars. Mergers do not appear to play a defining role in driving transitions between the three phases. The condition for thick disk formation appears to be minimal: a stable center of mass motion. Thin disk formation requires more: gas must accrete slowly enough for its angular momentum to mix and become coherent prior to joining the galaxy.

\end{abstract}




\section{Introduction}
Understanding the Milky Way's formation and dynamical evolution has been a core objective in astrophysics for well over half a century \citep[][]{ELS62}.  In addition to wanting to understand the Milky Way in its own right, there is hope that the richness of our understanding of our home galaxy will allow us to use it as a Rosetta Stone for general processes of galaxy formation. 
Indeed, wide-field spectroscopic surveys like SDSS, SEGUE, APOGEE, RAVE, GALAH, H3, and Gaia/Gaia-ESO provide detailed insights into the structure and makeup of the Galaxy not possible in  external systems \citep{Yanny2009,Gaia2016,Majewski2017,Conroy2019,Steinmetz2020,Buder2021}.

Among the more intriguing recent revelations about the Milky Way's past made possible by Gaia XP and APOGEE data \citep{aurora, Chandra2024} is that the in situ population of Milky Way stars displays three distinct kinematic phases. These phases are identified archaeologically by combining stellar phase-space information from Gaia with chemical information from APOGEE abundances and Gaia XP metallicities, and then examining how orbital structure changes as a function of metallicity. Specifically, \citet{aurora} and \citet{Chandra2024} showed that the evolution of stellar angular momentum or orbital circularity as a function of stellar metallicity or iron abundance (used as an approximate proxy for stellar age) transitions from 1) a disordered protogalaxy~\footnote{\citet{aurora} referred to the old isotropic component of in situ stars as ``Aurora.'' In this paper we will simply refer to this component as the protogalaxy or early Milky Way main progenitor.}, to 2) a kinematically hot older thick disk, to 3) a kinematically cold younger thin disk. The transitions between the phases are relatively sharp, with a characteristic ``spin-up'' metallicity, above which stars display coherent rotation, and a secondary ``cool down'' metallicity, above which the stars have thin disk kinematics and low dispersion.

\begin{table*}
  \centering 

  \begin{tabularx}{\textwidth}{Xcccccccccc}
    \hline
    \hline  
    Simulated Galaxy & $M_{\star}$ & ${R_\mathrm{90, gas}}$ & ${R_\mathrm{90, \bigstar}}$ & $M_{\mathrm{halo}}$ & $R_{\mathrm{halo}}$ & $m_{\mathrm{i}}$ & $t_{\mathrm{spin-up, gas}}$ & $t_{\mathrm{spin-up, \bigstar}}$ & $t_{\mathrm{CD}}$ & Reference \\
    Name  &  $[M_{\odot}]$ & [kpc] & [kpc] & $[M_{\odot}]$ & [kpc] & $[M_{\odot}]$ & [Gyr] & [Gyr]  & [Gyr] \\
    \hline\\[-0.32cm]
    Romeo     & $7.36\times 10^{10}$ & 18.56 & 13.29 & $1.24 \times 10^{12}$ & 282.5 & 3500 & 12.3  & 11.7  & 5.92  & A \\
    Juliet    & $4.22\times 10^{10}$ & 18.83 & 9.57 & $1.01 \times 10^{12}$ & 263.64 & 3500 & 8.96  & 7.75  & 3.46  & A \\
    Romulus   & $1.02\times 10^{11}$ & 18.47 & 14.21 & $1.88 \times 10^{12}$ & 321.69 & 4000 & 8.67  & 8.38  & 4.56  & D \\
    Remus     & $5.09\times 10^{10}$ & 18.44 & 12.28 & $1.13\times 10^{12}$ & 271.1 & 4000 & 11.66 & 9.35  & 5.30  & D \\
    Thelma    & $7.92\times 10^{10}$ & 17.97 & 12.44 & $1.32\times 10^{12}$ & 285.58 & 4000 & 7.8   & 4.6   & 2.35  & A \\
    Louise    & $2.85\times 10^ {10}$ & 18.75 & 12.05 & $1.03 \times 10^{12}$ & 263.09 & 4000 & 8.1   & 7.8   & 5.25  & A \\
    \hline\\[-0.31cm]
    {\tt m12b} & $9.42\times 10^{10}$ & 15.51 & 10.94 & $1.31\times 10^{12}$ & 286.18 & 7100 & 10    & 7.6   & 6.67  & A \\
    {\tt m12c} & $6.45\times 10^{10}$ & 16.02 & 10.37 & $1.26\times 10^{12}$ & 283.09 & 7100 & 7.6   & 6.5   & 2.81  & A \\
    {\tt m12f} & $8.78\times 10^{10}$ & 18.07 & 13.26 & $1.54\times 10^{12}$ & 302.1 & 7100 & 10.07 & 9.2   & 4.53  & B \\
    {\tt m12i} & $7.00\times 10^{10}$ & 17.13 & 9.98 & $1.07\times 10^{12}$ & 268.04 & 7100 & 7.67  & 7.32  & 3     & C \\
    {\tt m12m} & $1.26\times 10^{11}$ & 15.53 & 12.53 & $1.45\times 10^{12}$ & 296.43 & 7100 & 11.32 & 10.94 & 2.56  & E \\
    {\tt m12r} & $1.88\times 10^{10}$ & 17.24 & 13.01 & $1.03\times 10^{12}$ & 265.74 & 7100 & 7.03   & 6.5  & 4.78  & F \\
    \hline\\[-0.32cm]
  \end{tabularx}
  \raggedright
    \textit{Note}: The references are: 
	A:~\cite{Garrison-Kimmel19},
	B:~\cite{Garrison-Kimmel17}, 
	C:~\cite{Wetzel16},
	D:~\cite{Garrison-Kimmel19_2},
	E:~\cite{Hopkins18}, and 
	F:~\cite{Samuel20}.

  \caption{We summarize several key parameters of the simulations we use in our analysis. We include the following: $z = 0$ total stellar mass within 20 kpc of the host center, the radius that encloses 90\% of the $z = 0$ gas mass within 20 kpc as $R_{\rm 90, gas}$, the radius that encloses 90\% of the $z = 0$ young ($\leq 250$ Myr old) stellar mass within 20 kpc as $R_{\rm 90, \bigstar}$, the virial mass $M_{\rm halo}$ of the halo using the \protect\cite{bryan-and-norman} definition, the halo radius $R_{\rm halo}$ from the \protect\cite{bryan-and-norman} definition, the baryonic mass resolution $m_{\rm i}$, the gas spin-up time for each galaxy $t_{\mathrm{spin-up, gas}}$, the star spin-up time for each galaxy $t_{\mathrm{spin-up, \bigstar}}$ and the cooldown time for each galaxy $t_{\mathrm{CD}}$ (this should not be confused with $t_{\rm cool}$, which is the gas cooling time discussed later). The galaxies named `{\tt m12}' are isolated MW-mass halos, while the rest are pairs with Local Group-like environments from the ELVIS suite. We separate the ELVIS runs from the isolated {\tt m12} runs with a horizontal line. We also provide references to each halo. We define spin-up times for gas and stars as the times when the median circularities first rise in a sustained way above zero. We define the cooldown time as the time when the median circularity of the young stars approaches unity and when the dispersion is small, such that the 90th-percentile range remains above $\epsilon = 0.85$. Please see the end of Section \ref{sec:sample-results} for more discussion.}
\label{tab:one}
\end{table*}

These findings are particularly interesting in light of a series of papers pointing to a similar set of time-ordered phases seen in FIRE-2 simulations of Milky Way (MW)-mass galaxies. For example, \citet{Yu2021} showed that such galaxies  display sharp transitions from thick stellar disk formation at early/intermediate times to thin stellar disk formation at late times; this  transition coincides with a switch from bursty to steady star formation and the virialization of the inner circumgalactic medium \citep[][]{Stern_2021,bornthisway, Byrne2023, Sultan2026, Hafen_2022}.~\footnote{We note that such a transition is qualitatively consistent with the long-predicted trend that stellar disk formation proceeds from thick to thin \citep[e.g.][]{Brook04,Bird13,Ma2017,Bird20,Park20}.}  Separately, 
\citet{Gurvich23} looked at gas kinematics in FIRE-2 Milky Way progenitors, and in addition to quantifying the abruptness of the thick-to-thin transition, they found that in the earliest (pre-disk) stages,  the interstellar medium (ISM) of proto-Milky Ways are quasi-spheroidal with no clear boundary between the ISM and the inner circumgalactic medium (CGM). Looking again at stars, \citet{bornthisway} found that in situ populations in simulated Milky Way galaxies are naturally classified into three distinct components: 1) an old isotropic spheroid, 2) an intermediate-age thick disk, and 3) a young thin disk.  Importantly, they found that the orbital properties of stars at $z=0$ match well their orbital properties at birth.  That is, archaeological kinematic classifications of the kind explored by \citet{aurora} and \citet{Chandra2024} should reveal much about the structure of the galaxy at the time those stars were forming. 

Complementing this, \citet{McCluskey2024} analyzed stellar kinematics both at formation and at $z = 0$ using FIRE-2 MW-mass simulations to identify three kinematic eras (pre-disk, early disk, and late disk phases) that  parallel the protogalaxy, thick disk, and thin disk nomenclature for the phases discussed here. Across their sample, they found that the present-day stellar velocity dispersion is primarily inherited at birth rather than generated by subsequent dynamical heating for all but the oldest stars, implying that the archaeological dispersion structure of in situ stars largely retains memory of the galaxy's dynamical state at the time of formation. This further supports the approach of \citet{Chandra2024}.

The picture emerging from stellar archaeological evidence in the Milky Way is broadly consistent with what has been collected statistically from deep-sky surveys.  There is now strong evidence that galaxy disk populations evolve significantly over cosmic time \citep{Elmegreen07,Shapiro2008,Overzier2010,Elmegreen17}. 
Although most large star-forming galaxies in the local Universe have thin disks embedded in older thick disk populations, most star-forming galaxies at higher redshift have irregular morphologies that are dominated by massive star-forming regions with bright blue colors, typically called ``clumps'' \citep{Wuyts11,Zanella21}, whose formation and physical properties are also predicted by models of disk gravitational instability in turbulent, gas-rich systems \citep{Romeo_2010, Romeo_2014}. 

Many of these early disks are perturbed, thick, and turbulent, with lower rotation-to-dispersion ratios than local spirals.  Only at low redshift ($z \lesssim 1$) and in relatively massive galaxies does star formation begin to occur {\em primarily} in extended {\em thin} disks; this is sometimes referred to as the era of ``disk settling''  \citep{Kassin12,Wisnioski15,Tiley21}. Importantly, in a population sense, these trends are not sharp. Even out to $z \sim 8$, rotational support in cold gas has been seen to be quite significant, $v/ \sigma \sim 10$ in some systems \citep[e.g.,][]{Rowland24}, though with large uncertainties and in a limited number of cases.   More typically, at $z \sim 4-7$, thick disk-like rotational support with $v/ \sigma \sim 1 - 2$ is the norm \citep{Danhaive25}.
In related work, several studies with JWST have reported a surprisingly large fraction of systems with disk-like morphology at $z>2$ \citep[e.g.,][]{Ferreira2022,Robertson2023,Kartaltepe2023,Smethurst25}. However,  JWST is biased towards measuring diskiness in the brightest/most luminous systems at high-z, so the picture is still emerging. 

Stellar archaeology in nearby disks now provides a complementary view of how populations evolve. Resolved age–velocity dispersion measurements can trace the transition from turbulent early disks to dynamically cold thin disks on a galaxy-by-galaxy basis \citep{McCluskey2025}.

As observational evidence for the existence of these three phases grows, it is interesting to ask about the physics that drives the transitions from one to the next. This is the primary focus of our paper, which builds upon several past efforts. For example, \citet{Stern_2021} emphasized that the transition from cold-mode to hot-mode, or sub-sonic, gas accretion is important for understanding the thick-to-thin disk transition. \citet{Hafen_2022} showed that the ability of hot CGM gas to become coherent and angular-momentum supported prior to accretion onto the galaxy is key to forming thin, rather than thick, disks. \citet{2024MNRAS.532.3808M} argued that rapid gas cooling and rapid halo mass growth, as expected at high redshift, can cause gas to become self-gravitating before it settles into a disk, promoting spheroidal distributions at early times. This scenario has recently been tested by \citet{Ma2026} using FIRE-2 zoom-in simulations of MW-mass galaxies. They find that transitions in galaxy morphology, dynamical hotness, and star-formation burstiness correlate with the epoch when the host halo shifts from fast to slow accretion, suggesting that the two-phase assembly of halos may help regulate the transition from dynamically hot, bursty systems to colder, disk-dominated galaxies.

Another physical factor that can promote disk formation is the concentration of the gravitational potential, a point explicitly demonstrated by \citet{Hopkins_2023} using a series of numerical experiments.  Central mass concentrations provide well-defined dynamical centers, promote orbit mixing, stabilize disks, and suppress feedback-driven breathing modes.  

Feedback and outflows, in general, have long been understood as key to the successful formation of disk galaxies \citep[e.g.][]{Maller2002,Okamoto2005,Governato2007,Governato2010}.  One aspect explored by \citet{DeFelippis2017} is that initially low-angular-momentum gas in galaxies can gain angular momentum during wind ejections out in the halo and subsequently rain back in to contribute to disk formation. At $z = 0$, the fraction of specific angular momentum retained by the baryons correlates closely with disk morphology and bar structure across the present-day disk galaxy population \citep{Romeo_2022}. In a comprehensive analysis using TNG50 simulations, \citet{vadim1,vadim2} showed that corotating outflow recycling, gravitational potential steepening, and hot halo formation all correlate with disk formation, but were unable to definitively show causation. 

Galaxy gas fraction is another potentially important ingredient that affects the stability and formation of disks.  For example, \cite{Fensch21} have used a series of simulations to show that gas-rich models (which mimic high redshift expectations) form long-lived clumps and are prone to violent disk instabilities, whereas gas-poor models do not.  In contrast, gas-rich systems are more likely to re-form disks after major mergers \citep{2006ApJ...645..986R, Hopkins2009, GOvernato2009, Kannan15, 2017A&A...600A..25R}, and it may be the case that mergers of this kind promote disk formation at high redshift. In fact, the high-$\alpha$ chemistry of the Milky Way's in situ halo population suggests that the Gaia-Sausage-Enceladus (GSE) merger occurred after the formation of the early high-$\alpha$ disk and prior to the Galaxy's final cooldown into a thin disk \citep{Chandra2024, Belokurov2020, Bonaca2020}.
At the same time, major mergers involving galaxies with more modest gas fractions can disrupt or destroy disks \citep{1993ApJ...403...74Q, 2005A&A...438..507B} and potentially transform their morphologies into spheroids or ellipticals \citep{1972ApJ...178..623T, 2015ApJ...803...78T, 2007A&A...476.1179B}.

\begin{figure*}
    \centering
\includegraphics[width=\textwidth]{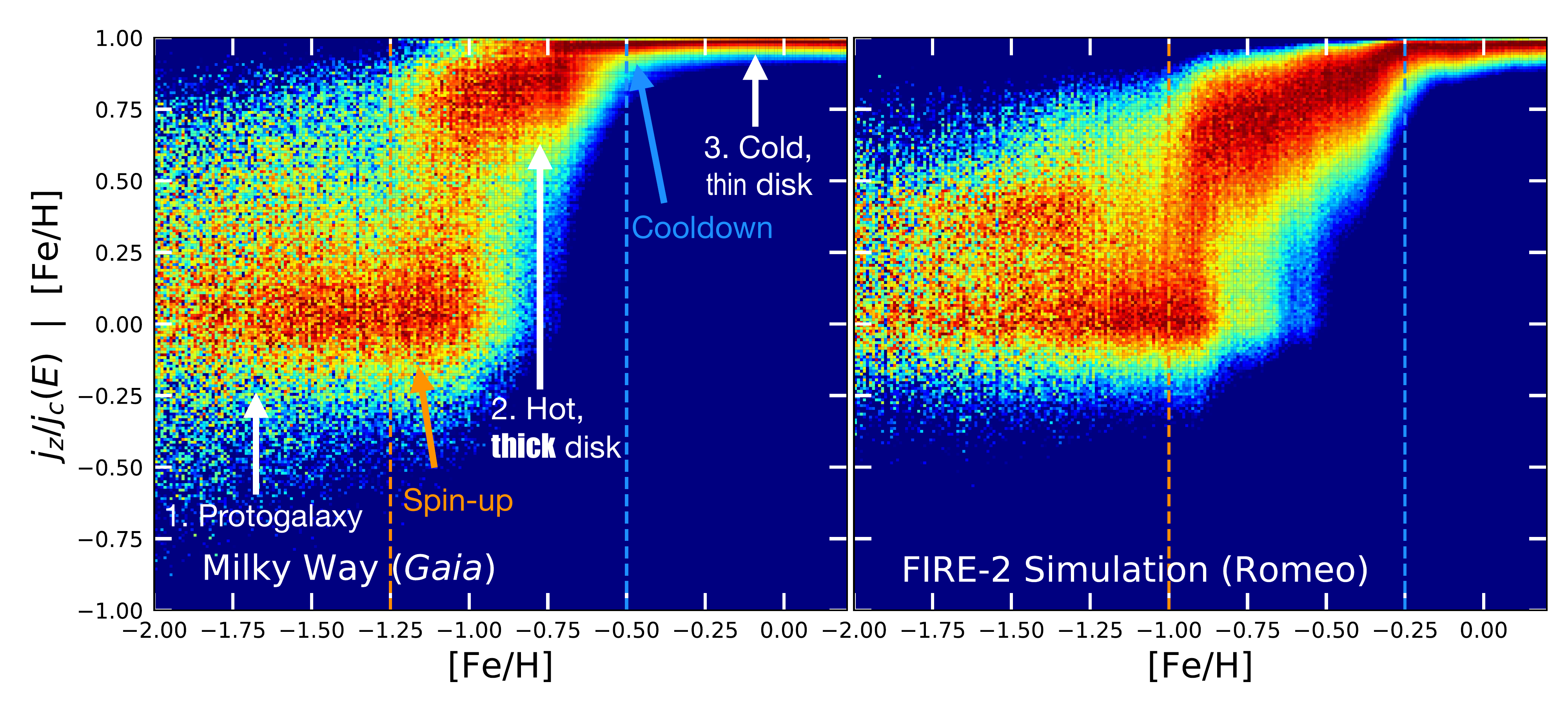}
    \caption{The three-phase evolution of Milky Way-size galaxies as revealed by observed Gaia RGB stars in the Galaxy (left) and the star particles in our Romeo simulation (right). The left panel reproduces Figure 5 from \citet{Chandra2024} and shows the column-normalized distribution of orbital circularities of $z = 0$ stars as a function of [Fe/H], which is a rough proxy for stellar formation time. Each column represents the distribution in stellar circularity at fixed [Fe/H], normalized to the maximum count along each respective column. Regions that are more red imply a greater count of stars with that circularity. The arrows and labels highlight three distinct kinematic phases of the Galaxy: 1) a disordered, isotropic protogalaxy, which subsequently spins up to 2) a kinematically hot, thick disk, and then cools down to 3) a thin disk. The right panel shows the same information for star particles within 20 kpc of the Romeo simulation. Note that we see the same three qualitative phases in the simulation as seen in the Milky Way, although the metallicities of the transitions are shifted higher ($\sim$ later times) in Romeo.}
    \label{fig:gaia and romeo three phases}
\end{figure*}

In the following sections, we use a suite of 12 MW-mass cosmological zoom-in simulations to explore the kinematic properties of stars and gas in the main progenitors over time. 
In Section \ref{sec:data-methods} we provide an overview of the data we use and our simulations. In Section~\ref{sec:example} we show an example simulation result as a way of demonstrating similarity to the Milky Way and defining timescales that mark transitions among the phases.  In Section \ref{sec:sample-results} we present sample-wide results.  Section \ref{sec:empirical-trends} is devoted to presenting empirical trends between kinematic phase, mass growth, and star formation. We then explore the physical origins of the kinematic phases in Section \ref{sec:causation}. Section \ref{sec:conclusion} summarizes our findings and presents some conclusions.

\par

\section{Data and Simulations}
\label{sec:data-methods}

\subsection{Gaia Data}
\label{sec:gaia-data}
We use an all-sky dataset of $\approx 10$~million red giant branch stars with spectrophotometry and radial velocities from the third data release of the \textit{Gaia} space observatory \citep{GaiaCollaboration2022,DeAngeli2022,Montegriffo2022a,Katz2022}. 
The construction of this sample is fully described in \cite{Chandra2024}. 
Briefly, stellar metallicities are drawn from \cite{Andrae2023}, who used a data-driven model to infer metallicities for stars with low-resolution \textit{Gaia} `XP' prism spectra. 
Stars belonging to bound structures like globular clusters are excised. 
\textit{Gaia} provides the sky coordinates, parallaxes, proper motions, and radial velocities for all these stars. 
Prior-informed parallax-based distances are drawn from the catalog of \cite{Bailer-Jones2020}. 
With these 6D phase-space coordinates in hand, Galactocentric coordinates and specific azimuthal angular momenta $j_\mathrm{z}$ are computed. 
We adopt a right-handed Galactocentric frame with a solar position $\mathbf{X}_\odot = (-8.12, 0.00, 0.02)$~kpc, and solar velocity $\mathbf{V}_\odot = (12.9, 245.6, 7.8)$~$\mathrm{km\,s^{-1}}$ \citep{Reid2004,Drimmel2018,GravityCollaboration2018}, and transform coordinates with \texttt{astropy} \citep{AstropyCollaboration2013, AstropyCollaboration2018, AstropyCollaboration2022} and \texttt{gala} \citep{gala,adrian_price_whelan_2020_4159870}.

\begin{figure*}
    \centering
    \includegraphics[width = \textwidth]{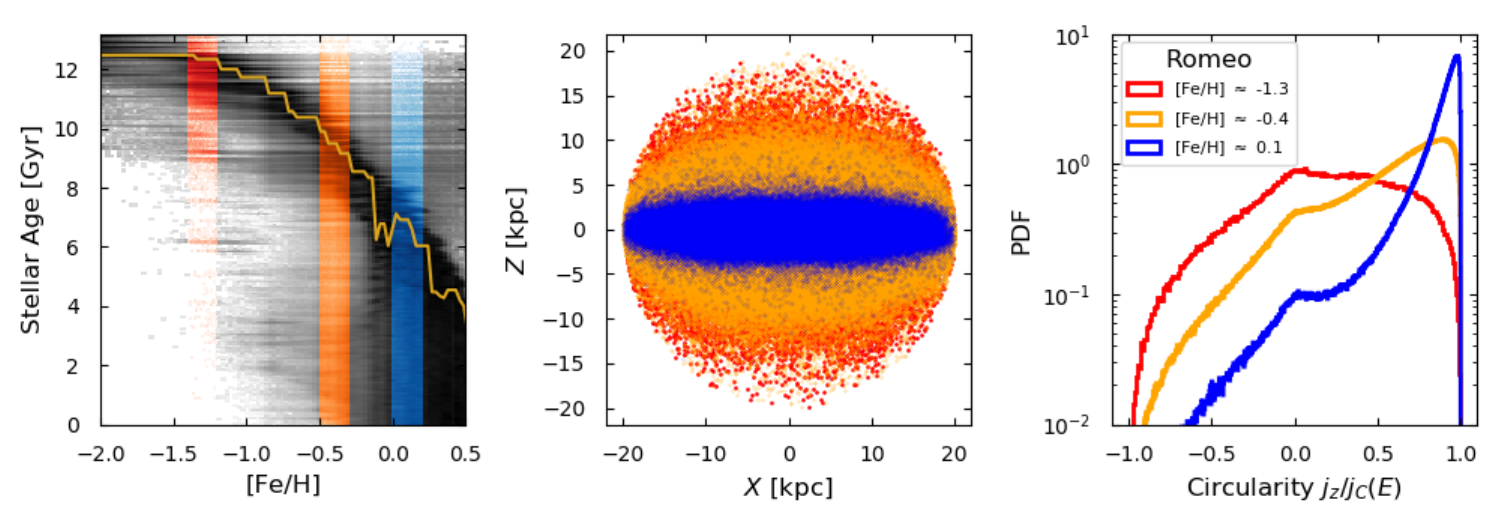}    \caption{Relationships between stellar age (measured in lookback time), [Fe/H], location in galaxy, and orbital circularity for present-day stars in the simulated  MW-mass galaxy Romeo. \textit{Left: } Column-normalized 2D histogram of stellar age at a fixed [Fe/H] versus [Fe/H] for all stars within 20 kpc of the galaxy center. The median of the 2D distribution is shown in yellow. The three colored bands pick out three groups of stars with different characteristic metallicities. \textit{Middle: } Spatial locations of the three groups of stars highlighted in the left panel. The youngest, most iron-rich stars (blue) primarily populate a disk, while the oldest, iron-poor stars (red) occupy a spheroidal distribution. Stars with intermediate ages and iron abundances resemble a flattened spheroid or a very thick disk. \textit{Right: } Distribution of orbital circularities for each of these selected groups of stars. The most metal-poor stars have a fairly isotropic circularity distribution centered on $\epsilon = 0$. The intermediate-metallicity/age stars have a preference for co-rotation, while the youngest, most metal-rich group has an orbital distribution peaking near $\epsilon \sim 1$.}  
    \label{fig:jzjc pdf scatter}
\end{figure*}

\subsection{Simulation Data}
We use cosmological baryonic zoom-in simulations from the Feedback In Realistic Environments (FIRE) project, run using the multi-method gravity plus hydrodynamic code {\tt GIZMO} in its mesh-free finite mass (MFM) mode \citep{Hopkins15} \footnote{https://fire.northwestern.edu/} with FIRE-2 physics implementations \citep{Hopkins18}. Briefly, the simulations include a spatially uniform, meta-galactic UV/X-ray ionizing background \citep{2009ApJ...703.1416F}, and multi-channel stellar feedback including continuous mass loss (stellar winds) from OB and AGB stars, supernovae type Ia and II, photoelectric heating and radiation pressure. Stellar evolution models directly inform the input of the parameters defining these heating sources. Our simulations self-consistently generate and track the evolution of 11 elemental gas and stellar abundances and their sub-grid diffusion via turbulence: H, He, C, N, O, Ne, Mg, Si, S, Ca and Fe \citep{Hopkins16, Su2019, Escala2018}. Star formation requires  gas to be locally self-gravitating, dense ($> 1000 \, {\rm cm}^{-3}$),  Jeans unstable, and molecular \citep{Krumholz2011}. Local star formation efficiency is set to be 100\% per free-fall time, or ${\rm SFR}_{\rm particle} = m_{\rm particle} \cdot f_{\rm mol} / t_{\rm ff}$ such that gas particles are stochastically converted into stars at this rate \citep{Katz96}. This local condition, however, does {\it not} imply that the global efficiency of star formation within the scale of, say, a giant molecular cloud (GMC) or larger is 100\%. In fact, the self-regulation of feedback limits star formation to $\sim 1-10\%$ per free-fall time \citep{CAFG2013, Hopkins17, Orr18}. \par

Since stellar iron abundance plays a role in our preliminary comparison to observations below,  we note that FIRE-2 uses nucleosynthetic yields for stellar feedback from \cite{Nomoto2006} (Type II supernovae) and \cite{Iwamoto1999} (Type Ia supernovae). The rates for Type Ia supernovae are sourced from \cite{Mannucci2006}, but see \cite{Gandhi2022} for previous analysis examining the effects of different Type Ia supernova rates on stellar iron abundances of FIRE-2 galaxies. Note that \citet{Bellardini2022} showed that FIRE-2 galaxies exhibit a systematic median age–metallicity relation, albeit with substantial scatter.

Six of our galaxies with the naming convention `{\tt m12}$\bullet$' are isolated field MW-mass halos from the Latte simulation suite \citep{Wetzel16, Garrison-Kimmel17, Hopkins17, Garrison-Kimmel19}. We also use six galaxies that come in three pairs from the ELVIS project \citep{Garrison-Kimmel2014, Garrison-Kimmel19, Garrison-Kimmel19_2}. We analyze each of the constituent host galaxies of these three pairs separately, giving us a total of six MW-mass systems in our sample from the ELVIS project. These galaxies are members of Local Group-like environments selected such that the main halos have similar $z = 0$ relative separations and velocities as the MW/M31 Local Group system.  \par 

We summarize key properties of our simulated galaxies in Table \ref{tab:one}. These properties include: the $z = 0$ total stellar mass within 20 kpc ($M_\star$), radii enclosing 90\% ($R_{90}$) of gas and young $(\leq 250$ Myr old) stellar masses at $z = 0$, virial masses ($M_{\rm halo}$) and radii ($R_{\rm halo}$), initial baryonic particle mass resolution, and transition times between kinematic phases (described below).  \par 

\par

\section{Spin-up and Cooldown Times in Simulated Galaxies}
\label{sec:results}

We begin with a description of nomenclature. In the next subsection and throughout this paper, we adopt the terminology and naming conventions of the three kinematic phases introduced in \cite{Chandra2024}. Specifically, we use the term  \textit{spin-up} to refer to the transition between the disordered protogalaxy phase and the early thick disk stage. Importantly, spin-up does {\em not} mean that pre-existing stars are externally torqued up at this time. Rather, it marks a change in the types of orbits that {\em new} stars have. Prior to this time, new stars tended to be born on radial orbits.  After this time, they are born with more coherent spin. Similarly, we refer to the \textit{cooldown} time as the transition between the thick disk and thin disk. Note that during the thick disk phase, new stars often gradually start to occupy orbits with higher circularities, a process previously studied in FIRE-2 simulations of MW-mass galaxies \citep{Ma2017, McCluskey2024, Gurvich23}. This settling process is {\it not} what we are referring to as cooldown. Instead, we refer to cooldown as the final transition to the time when new stars are born exclusively on thin disk orbits.

\subsection{Example Case: Comparing Romeo to the Milky Way}
\label{sec:example}


In this subsection we use one of our simulated galaxies, Romeo, as an example case, before moving towards a sample-wide exploration in the next subsection. Of our 12 simulated galaxies, Romeo has among the earliest-forming thin disks.

The key result from \citet{Chandra2024} (that the MW evolved in three distinct kinematic phases) is shown in the left panel of Figure \ref{fig:gaia and romeo three phases}.  To be specific, we reproduce Figure 5 from \citet{Chandra2024} and present the column-normalized distribution of orbital circularity as a function of Gaia XP metallicity for their all-sky giant star sample. Circularity is defined as $\epsilon \equiv j_z/j_C(E)$ \citep{Abadi2003}, where $j_z$ is the specific angular momentum of the star, or particle, in the average direction of angular momentum of the system $z$, and $j_C(E)$ is the specific angular momentum of a circular orbit with the same specific orbital energy, $E=\frac{1}{2}|\mathbf{v}|^2+\Phi(r)$. For the simulations, we compute $\Phi(r)$, the gravitational potential per unit mass at radius $r$, from the spherically averaged cumulative mass profile of all stars, gas, and DM. The circular-orbit radius corresponding to each particle energy is then found in this same spherical potential and used to compute $j_C(E)$.  The value of $\epsilon$ can range from $-1$ to $1$. $\epsilon = 1$ implies a perfectly circular prograde orbit in the galaxy's plane of rotation, $\epsilon = 0$ implies a purely radial orbit, and $\epsilon = -1$ implies a perfectly retrograde orbit. The three kinematic phases are indicated in white in Figure \ref{fig:gaia and romeo three phases}: 1) Protogalaxy; 2) Hot, thick disk; and 3) Cold, thin disk.

On the right in Figure \ref{fig:gaia and romeo three phases} we show the same information for the Romeo simulation.  The only difference is that we show $z=0$ star particles within 20 kpc of the galaxy center rather than individual giant stars from their all-sky survey.  In both the observed Milky Way and the simulation, we notice three clear kinematic phases in evolution: an early, chaotic protogalaxy with little to no net rotation, a kinematically hot thick disk, and a kinematically cold thin disk. The ``spin-up'' transition from protogalaxy to thick disk is marked by a vertical orange line in each panel.  The ``cooldown'' transition from thick disk to thin disk is marked by the blue vertical line in each panel. We note that the Milky Way transition metallicities are systematically more metal-poor than those seen in the Romeo simulation; this suggests that the Milky Way's thick and thin disks both formed earlier than their counterparts in Romeo.  \footnote{In both panels of Figure \ref{fig:gaia and romeo three phases} we see that while most $j_z/j_c$ values scatter about $\sim 0$, there is a minority population with some positive spin at [Fe/H] $\approx -1.3$. This phenomenon has previously been identified in the MW by \cite{aurora} and has been dubbed `Aurora', a kinematically hot, approximately isotropic velocity ellipsoid possessing a weak net rotation.} 

A natural question raised by this decomposition is how much stellar mass is associated with each of the three phases. We return to this point in Appendix~\ref{sec:phase_masses}, where we quantify the net stellar mass growth within 20 kpc during the protogalaxy, thick disk, and thin disk phases for Romeo and for the full simulation suite. For Romeo, most of the present-day stellar mass is accumulated during the thick disk phase, although the corresponding phase mass fractions vary substantially across the suite.

\begin{figure*}
    \centering 
    \includegraphics[width=1 \textwidth]{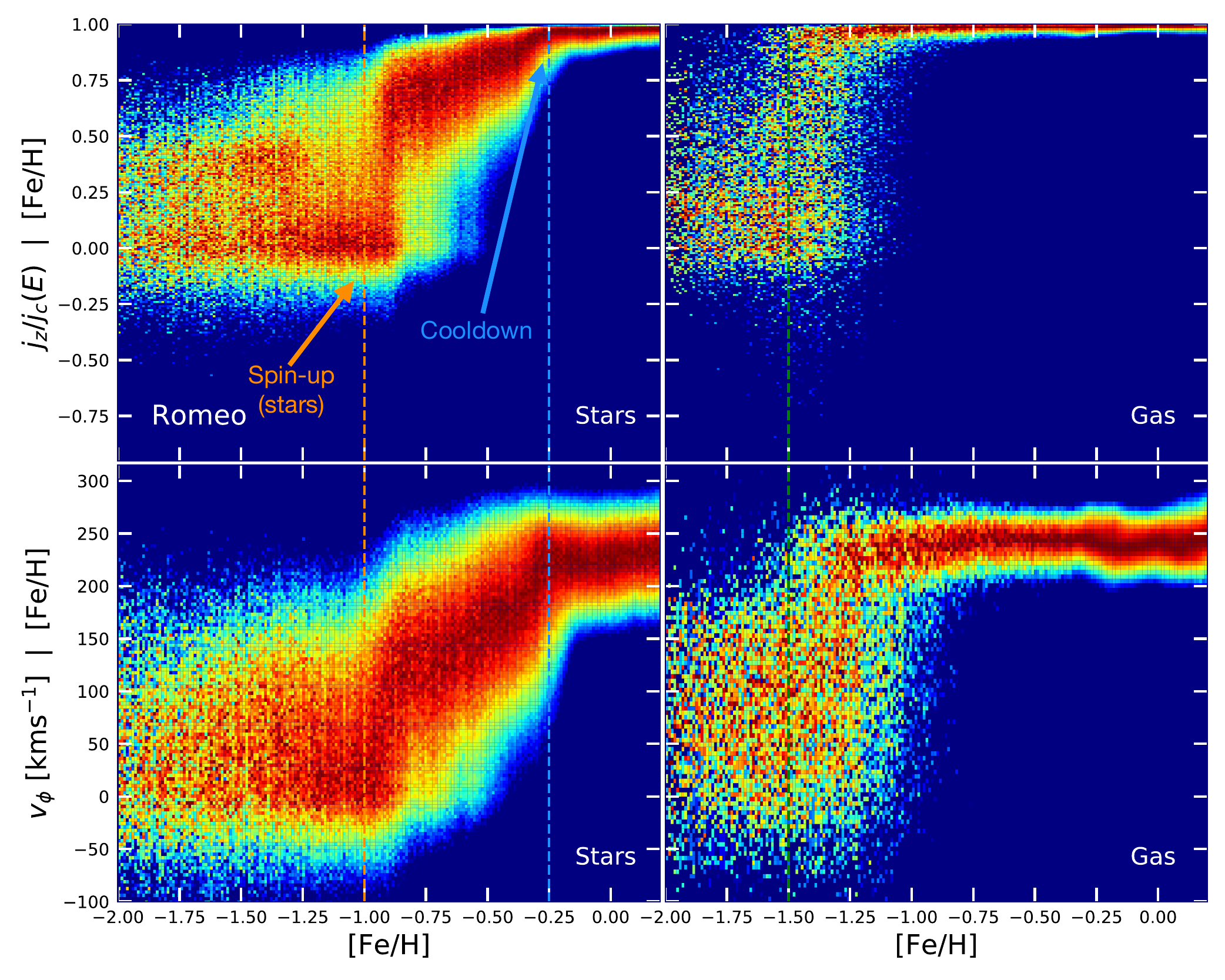}
    \caption[]{Column-normalized 2D histograms of $z=0$ orbital properties as a function of metallicity in Romeo. The left panels show star particles within 20 kpc of the center; the right panels show gas particles in the same region. The top panels show orbital circularity ($j_z/j_c(E)$) and the bottom panels show tangential velocity ($v_\phi$). The stars display all three kinematic phases (i.e. protogalaxy, thick disk, and thin disk), while the gas appears to show only two broad components in this parameter space, with no clearly separated intermediate phase. This difference likely reflects the fact that we plot present-day gas properties versus $[{\rm Fe/H}]$ rather than versus time: in time-based analyses, gas rotational support grows gradually prior to cooldown \citep[e.g.,][]{Gurvich23}. At fixed $[{\rm Fe/H}]$, however, the bursty pre-cooldown era is associated with strong inflows, outflows, and mixing between the disk and inner CGM \citep{Stern_2021, Gurvich23}, which distribute gas of similar metallicity across a wide range of $v_\phi$ and $j_z/j_c$ and effectively smear out an intermediate phase. By cooldown, when outflows are strongly suppressed, the gas settles into a narrow, high-circularity configuration at high $[{\rm Fe/H}]$. Stars, by contrast, retain their birth metallicities and orbital structure more permanently, and therefore preserve the full three-phase picture more clearly in this parameter space. We also note that the rotationally supported $z=0$ gas component extends to lower [Fe/H] than the stellar spin-up feature, and that the gas disk occupies a noticeably narrower band in circularity than the stellar disk, as expected since gas can dissipate energy while stars cannot.
    
    
    }
\label{fig:jzjc_vphi_feh_stars_gas}
\end{figure*}

In identifying these three kinematic phases using \textit{Gaia} data, \cite{Chandra2024} use a select sample of red giant branch stars, described in Section~\ref{sec:gaia-data}. In contrast, we show this three-phase evolution in Figures \ref{fig:gaia and romeo three phases} and \ref{fig:jzjc_vphi_feh_stars_gas} with simulation data using {\it all} the stars within 20 kpc of the host center. For the purpose of illustrating the qualitative resemblance of three-phase kinematic evolution between Romeo and the Milky Way, we do not use a sample of stars in our simulation data that directly mimics \cite{Chandra2024}'s {\it Gaia} sample. While different selection criteria in our simulation data can alter minor features in the right-hand panel of Figure \ref{fig:gaia and romeo three phases}, regardless of these criteria, the three-phase kinematic evolution of Romeo visually dominates in all cases. For example, even if we generate a more observationally motivated sample instead (as we show in Figure~\ref{fig:z_R_jzjc_fe_over_h_romeo} and discuss in Appendix~\ref{sec:selection}), we still find a striking qualitative result of three distinct phases. Generally speaking, we are not concerned with producing an exactly analogous stellar sample that compares with \cite{Chandra2024}, but instead with exploring and reporting on the gross physical processes associated with disk formation in MW-mass systems. Future studies with this same simulation dataset, however, can be used to provide a more careful comparison with the \textit{Gaia} data used in \cite{Chandra2024}.

\begin{figure}
    \centering 
    \includegraphics[width=1 \columnwidth]{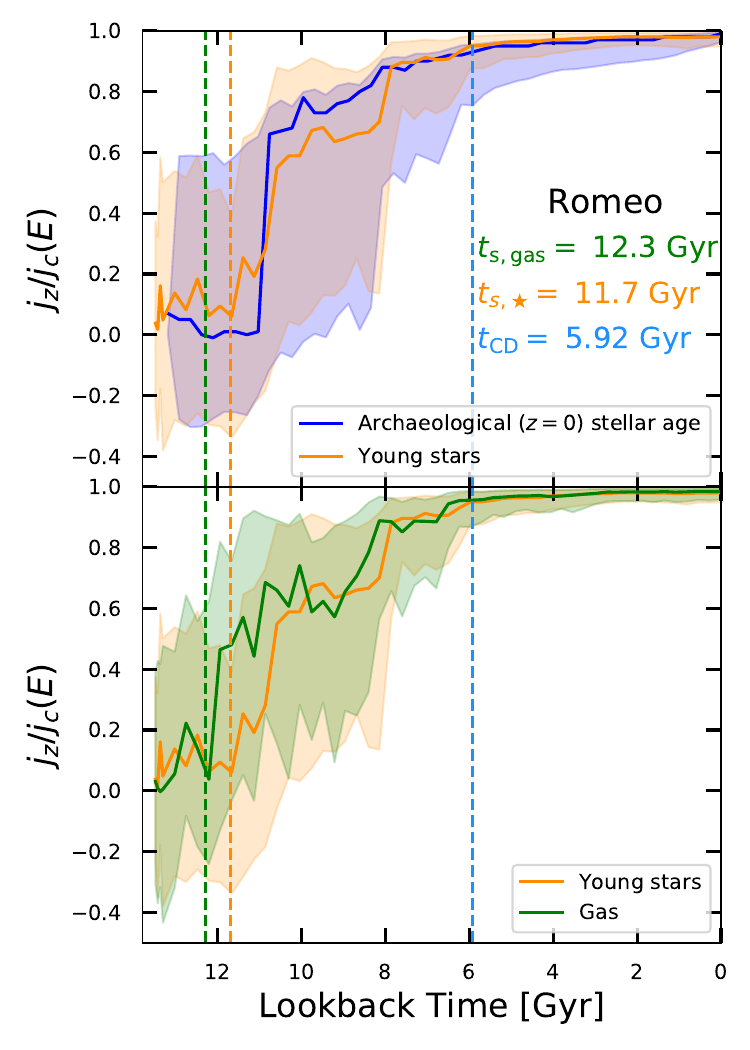}
    \caption[]{{\bf Top:} The blue shaded band shows the ``archaeological'' orbital circularity distributions of all stars within 20 kpc of Romeo at $z=0$ as a function of their stellar age.  The median circularity is depicted as a solid blue line.  The shading shows the 90th-percentile distribution.  The orange band and line show the same information for young stars ($<250$ Myr old) as a function of lookback time when those stars were forming in the galaxy. The vertical dashed orange line shows the ``stellar spin-up time'' (11.7 Gyr ago) and the vertical dashed blue line shows the ``cooldown time'' (6 Gyr ago). We see that the behavior as a function of stellar age and as a function of lookback time are similar, except that the archaeological method shows a slightly higher dispersion after cooldown, a result of kinematic heating.  {\bf Bottom:} The young star distribution is reproduced from the top panel in orange along with the circularity distribution of all gas within 20 kpc in green.  As indicated by the vertical green dashed line, the gas spins up about 500 Myr earlier than the young stars, $\sim 12.3$ Gyr ago.  The gas and stars have the same cooldown time.}
    \label{fig:romeo median and 1sigma}
\end{figure}

We note that neither panel of Figure~\ref{fig:gaia and romeo three phases} separates in situ from accreted stars. At low metallicity, Gaia--Sausage--Enceladus debris can sharpen the apparent Milky Way spin-up transition by adding stars with little net angular momentum. Correspondingly, the predominantly in situ high-$\alpha$ population shows a more gradual spin-up than the low-$\alpha$ population (\citealt{Chandra2024}; \citealt{Viswanathan2025}). Our simulation panel likewise includes all stars within 20 kpc, but the bulk of our subsequent analysis follows young stars in the main progenitor as a function of lookback time, which closely tracks the archaeological measurement in the top panel of Figure~\ref{fig:romeo median and 1sigma}.


Figure \ref{fig:jzjc pdf scatter} provides an illustrative example of how the observable parameters in Figure \ref{fig:gaia and romeo three phases} map to age and spatial structures in Romeo.  The left panel of Figure \ref{fig:jzjc pdf scatter} shows the column-normalized relationship between $z=0$ star particle age and star particle metallicity [Fe/H]. The yellow line shows the median age at fixed [Fe/H], demonstrating that the most metal-rich stars are typically younger than metal-poor stars.  The middle panel of Figure \ref{fig:jzjc pdf scatter} shows the spatial location of three sets of star particles that sample the metallicity distribution as colored vertical bands in the left panel. Here, the vertical (z) component is defined as the direction of net angular momentum of the stars. We see that the highest metallicity band (blue) resides in a disk-shaped component. The modestly metal-poor sample (orange) is arranged in a  thick disk configuration, and the most metal-poor sample (red) is quite isotropically distributed.

The right panel of Figure \ref{fig:jzjc pdf scatter} shows the circularity distributions for those same sets of stars with the same color code. We see that the metal-rich sample (blue) has a circularity distribution that is peaked towards $\epsilon \sim 1$, indicative of a prograde disk.  The most metal-poor sample (red) has a wide range of circularities, indicative of a kinematically isotropic sample. The moderate metallicity sample shows a slight preference for co-rotation. Comparing the circularity histograms shown in the right panel of Figure \ref{fig:jzjc pdf scatter} to the spatial locations of particles in the middle panel also provides an intuitive guide to how stellar circularity relates to spatial structure. For example, the young/metal-rich population (blue), with prograde, disk-like orbits in the right panel, does indeed inhabit a spatial disk in the middle panel.

The upper left panel of Figure \ref{fig:jzjc_vphi_feh_stars_gas} reproduces the right panel of Figure \ref{fig:gaia and romeo three phases}, and the lower left panel shows column-normalized tangential velocities as a function of metallicity; this figure can be compared to Figure 4 in \citet[][]{aurora}. The right panels of Figure \ref{fig:jzjc_vphi_feh_stars_gas} show the same information for all gas particles within the same volume. There is no temperature cut here. It is important to note that stellar and gas-phase metallicities encode different information in this figure: stellar $[{\rm Fe/H}]$ is set at the time of formation and is then largely preserved, whereas gas-phase $[{\rm Fe/H}]$ is an instantaneous property of the gas at $z=0$. The stellar panels therefore provide an archaeological record of the kinematics of stars born at different enrichment stages, while the gas panels show only the present-day relationship between gas metallicity and gas kinematics. The $z=0$ gas displays two clear kinematic phases: a thin disk at high metallicity and a kinematically broad spheroid/halo component at low metallicity, with no clearly separated intermediate phase. One difference is that the gas-disk feature appears at a lower metallicity ($[{\rm Fe/H}] \sim -1.5$, green lines on right) than the stellar spin-up metallicity ($[{\rm Fe/H}] \sim -1$, orange line on left). Moreover, the gas-disk feature becomes sharply focused around $[{\rm Fe/H}] \sim -1$, which is close to the spin-up metallicity for the stars. Finally, it is interesting to note that both the inner metal-poor gas halo and the most metal-poor ``protogalactic'' stars are mildly co-rotating ($V_c \sim 50$ km/s) in this simulation.


\citet{Viswanathan2025} find that, once the predominantly in situ high-$\alpha$ population is considered separately from the low-$\alpha$ population (whose metal-poor end contains a substantial contribution from accreted debris), the Milky Way's spin-up proceeds gradually with increasing metallicity in both $v_\phi$ and orbital circularity. Separately, Figure~\ref{fig:jzjc_vphi_feh_stars_gas} shows that in Romeo the stellar $v_\phi$-[Fe/H] relation rises more gradually than the corresponding circularity-[Fe/H] relation. This does not necessarily imply that the physical transition itself is gradual. In the gas panels, the emergence of a rotationally supported disk remains sharp in both $v_\phi$ and circularity. The smoother stellar $v_\phi$ trend likely reflects the fact that $v_\phi$ is a less direct archaeological diagnostic than circularity: it depends on present-day radius, the local circular speed, and subsequent heating or radial mixing. This means that populations with different spatial distributions can exhibit a broadened median $v_\phi$-[Fe/H] sequence, even if the transition in their birth orbital structure was comparatively abrupt. We therefore regard circularity as the cleaner diagnostic of spin-up in this context, while $v_\phi$ provides a useful but more smeared projection of the same underlying transition.

Of course, [Fe/H] is only a rough  proxy for stellar formation time. As shown in the left panel of Figure \ref{fig:jzjc pdf scatter}, the median stellar age in Romeo varies monotonically with [Fe/H] only down to [Fe/H]$\simeq -2$.  Moreover, there is substantial scatter in age at fixed metallicity throughout.  For this reason, we now shift to presenting results as a function of either stellar age or lookback time as measured in the simulations. The results that follow do not rely on converting [Fe/H] into age. \par

Figure \ref{fig:romeo median and 1sigma} presents a direct exploration of the kinematic phases seen in Romeo as a function of stellar age and lookback time. The blue solid line (shaded region) in the top panel shows the median (90th-percentile range of) stellar circularity as a function of stellar age.  This is analogous to archaeological explorations shown in Figures \ref{fig:jzjc_vphi_feh_stars_gas} and \ref{fig:gaia and romeo three phases}. The orange line and shading instead show the circularity of {\em young stars in the main progenitor} as a function of {\em lookback time}. Here, ``young stars'' refers to stars younger than 250 Myr at each timestep. The similarities between these two measures are notable.  Specifically, the archaeological picture is very much like the picture drawn from directly tracking the galaxy over time.  Both trajectories reveal an early isotropic protogalaxy, which transitions to a co-rotating disk with high dispersion, and finally a late-time thin disk.  The main difference is that the archaeological thin disk is slightly hotter, owing to kinematic heating that has occurred since the stars formed, which has also been explored extensively in \cite{McCluskey2024} across the full FIRE-2 suite. The two measurements need not agree in detail, however, as shown by \citet{Orkney_2026} using the Auriga simulations: spin-up times inferred from $z = 0 $ stellar kinematics are systematically delayed relative to those measured from birth kinematics, with the offset driven largely by massive radial mergers that erase the kinematic signature of a pre-existing disk.

The bottom row of Figure \ref{fig:romeo median and 1sigma} reproduces the young-star circularities from the top panel in orange and also shows the circularity distribution of gas particles in the main progenitor as a function of lookback time in green.  Notably, the gas in the galaxy follows a similar trajectory.  The only difference is that the gas spins up slightly earlier than the young stars.  Both the gas and young stars transition to the thin disk phase at about the same time.

\begin{figure*}
    \centering
    \includegraphics[width=\textwidth]{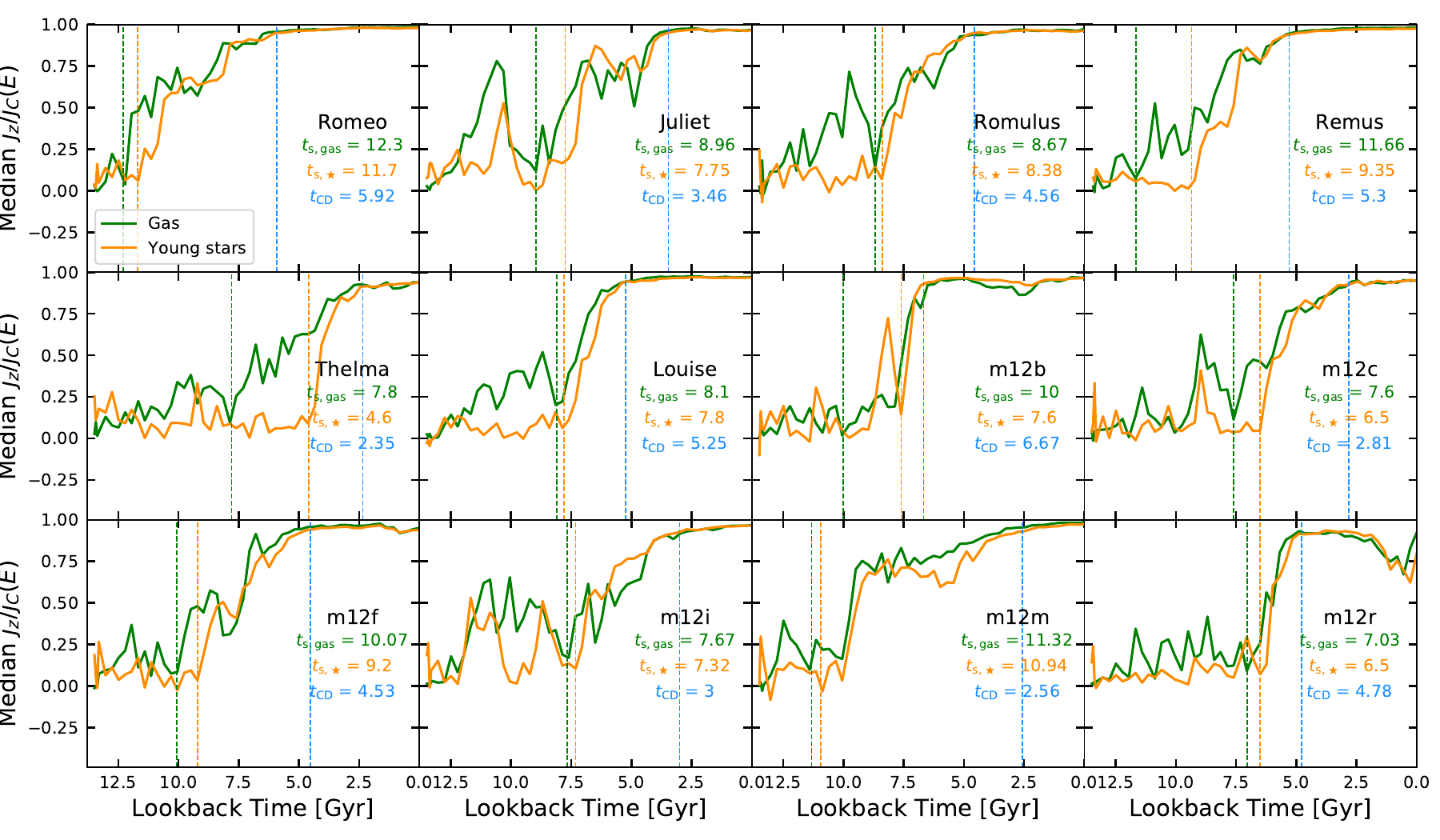}
    \caption{Median circularity as a function of lookback time for young stars (orange) and gas (green) for our full simulation suite. For each simulated galaxy, we indicate the gas spin-up time (green vertical line), stellar spin-up time (orange vertical line), and cooldown time (blue vertical line). All of the simulations display a similar series of kinematic phases.  Note that m12r (bottom right) experiences a late-time major merger, which drives down the median circularities after the cooldown phase.}
    \label{fig:fire suite jzjc medians}
\end{figure*}

\begin{figure*}
    \centering
\includegraphics[width=\textwidth]{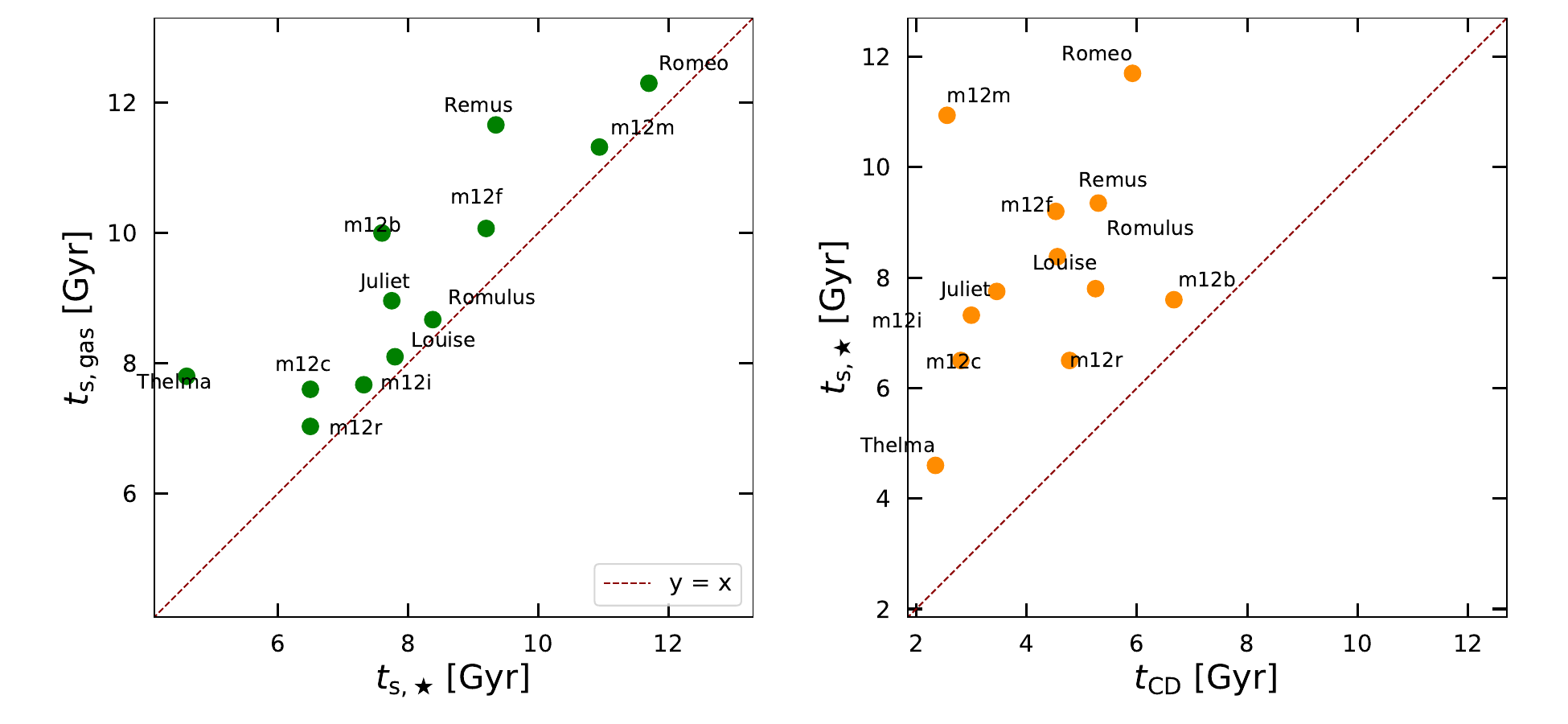}
    \caption{Gas spins up before stars and stars spin up before cooldown. {\bf Left:} We plot gas spin-up lookback times against stellar spin-up lookback times and see that gas always spins up first. Many systems have an extremely short delay between gas and stellar spin-ups. {\bf Right:} We plot stellar spin-up times against cooldown times and see that stars generally spin up well before our galaxies cool down.}
    \label{fig:t_gas t_star t_cool}
\end{figure*}

The three vertical lines in Figure \ref{fig:romeo median and 1sigma} introduce characteristic transition times that we will use in this paper going forward: 1) the gas spin-up time, $t_{\rm s, gas}$; 2) the stellar spin-up time, $t_{\rm s, \star}$, and 3) the cooldown time, $t_{\rm CD}$.  These times are based on the main progenitor (lookback time) trajectories. We define spin-up times for gas and stars as the times when the median circularities first rise in a sustained way above zero. We define the cooldown time as the time when the median circularity of the young stars approaches unity {\em and} when the dispersion is small, such that the 90th-percentile range remains above $\epsilon = 0.85$. We emphasize again that the gas spins up before the young stars spin up; as discussed below, this is a trend that we observe for the sample more generally.

Our definition of the spin-up time is empirical and sample-specific. For each of the 12 FIRE galaxies analyzed here, we use the time evolution of the median circularity distribution as the primary diagnostic to identify spin-up as the onset of a sustained rise away from the near-zero values associated with the protogalaxy phase. While this procedure involves judgment at the level of individual systems, it is not arbitrary: it is anchored to a quantitatively measured feature of the circularity evolution, applied consistently across the sample. We use it only as an operational definition for these 12 simulated galaxies and do not claim that it implies a universal threshold or a prescription that should be extrapolated beyond this dataset.\par

\begin{figure}
    \centering 
    \includegraphics[width = 1 \columnwidth]{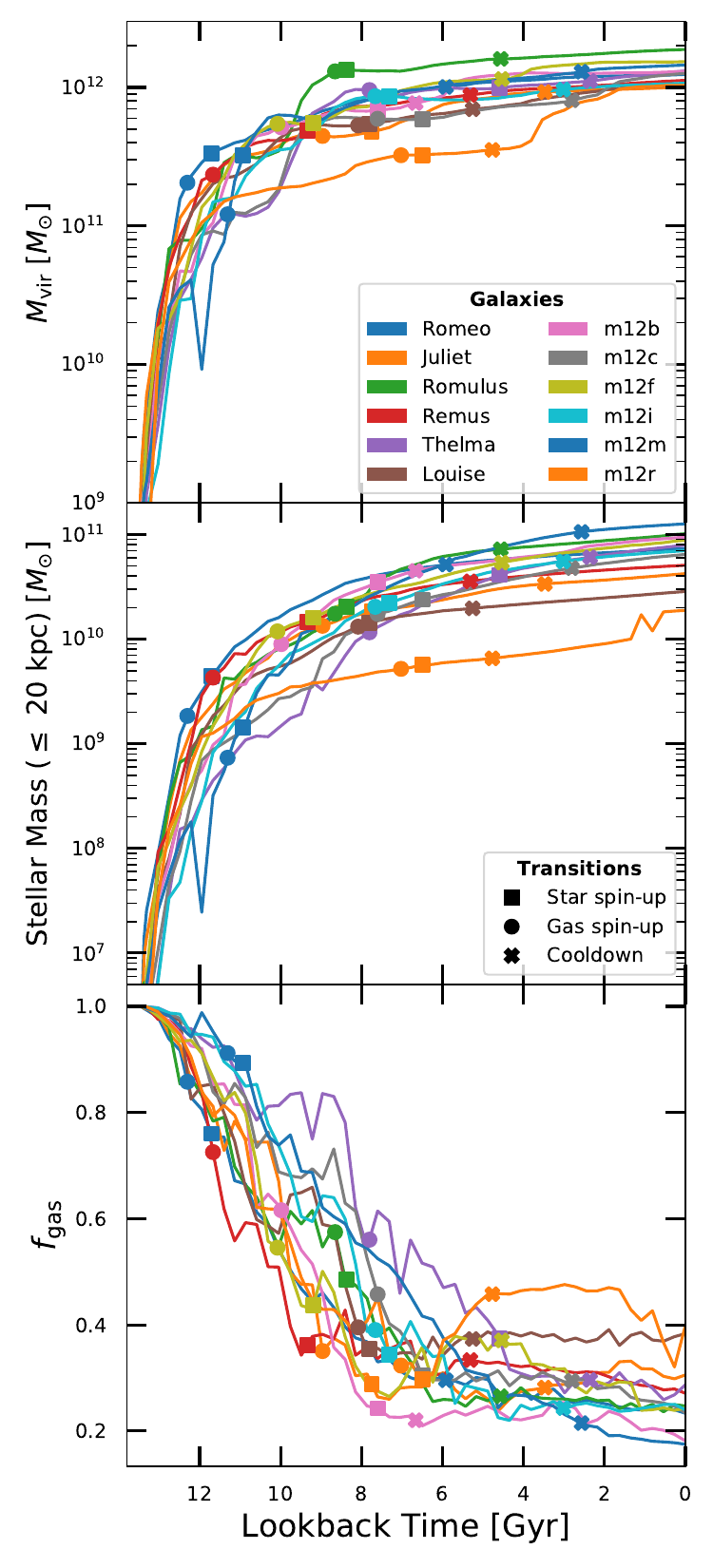}
    \caption[]{Virial mass (top), stellar mass (middle), and gas fraction (bottom) as functions of lookback time for each galaxy in our simulation suite. Note that $f_{\rm gas} = M_{\rm gas} / (M_{\rm gas} + M_{\bigstar})$ and is calculated within 20 kpc. We mark the value at the spin-up time for each galaxy. Given the large scatter in values of these parameters at each galaxy's spin-up time, there is no conclusive mass threshold to fully characterize spin-up in either stars or gas, but we find that stellar spin-up generally occurs across the range $M_{\rm vir} \sim 3-10\times 10^{11} M_\odot$, and $M_\bigstar \sim 3-10\times 10^{9} M_\odot$. We also show the maximum circular velocity of these galaxies as functions of lookback time in Appendix \ref{sec:v_max}.}
\label{fig:characteristic masses and gas fraction}
\end{figure}

\subsection{Sample-wide Results}
\label{sec:sample-results}

The twelve panels of Figure \ref{fig:fire suite jzjc medians} present circularity as a function of lookback time for each of our simulated galaxies. The green lines show the median circularity of all gas particles within 20 kpc of the main progenitor.  The orange lines show the median circularity of young stars (< 250 Myr) within the same region. The vertical lines mark the gas spin-up times (green), star spin-up times (orange), and cooldown times (blue) for each galaxy. Note that the gas tends to spin up slightly before the stars. We do not distinguish stellar cooldown and gas cooldown times because they are always almost simultaneous. 

Figure \ref{fig:fire suite jzjc medians} demonstrates that all galaxies in our simulation suite experience the three kinematic phases identified in the previous section. In all but one case, the median circularity trajectories become sharply focused at $\epsilon \simeq 1$ toward the present day. That is, they all ``cool down'' such that the young stars and gas enter a thin disk phase at late times.   The exception is m12r, which experienced a major merger about 1 Gyr prior to $z=0$. About 4 Gyr before that, m12r transitioned to its thin disk phase with median circularity near unity.  The circularity then dips to $\sim 0.75$ after the merger before beginning to rise again towards unity, with the gas spinning up again slightly before the young stars.

Another trend apparent in Figure \ref{fig:fire suite jzjc medians} is that every galaxy has an early phase when both young stars and gas have an isotropic orbit distribution with $\epsilon \sim 0$ prior to spin-up.  There are several cases (e.g., Juliet, Romulus, m12c, m12i) where the median circularity oscillates between some degree of rotation ($\epsilon \sim 0.5$) and isotropy prior to the sustained transition towards rotation. We identify the spin-up time to be the point after which the growth in circularity is monotonic, and we chose these times subjectively for each run by eye. 

Figure \ref{fig:fire suite jzjc medians} also shows that the gas tends to spin up before the young stars. In some cases, the difference is only a few hundred million years, consistent with the age range of ``young stars'' by our definition, but is as large as 2 Gyr in Remus, m12b, and Thelma (Table~\ref{tab:one}). That gas spins up more readily is perhaps expected: gas is collisional and can dissipate to angular momentum support, whereas stars cannot.
\par

The difference in the spin-up times for gas and stars may also reflect a difference between the broader gas reservoir and the cool, dense gas that is actively forming stars.  In particular, we find that star-forming gas in our clumpy and rapidly evolving galaxies is subject to strong gravitational torques that continually rearrange its angular momentum, even after the surrounding, lower density gas has begun to rotate coherently. Stars born during this period retain a memory of these disturbed gas motions, while the gas can dissipate energy, reorient, and settle again.


\begin{figure*}
    \centering 
    \includegraphics[width = \textwidth]{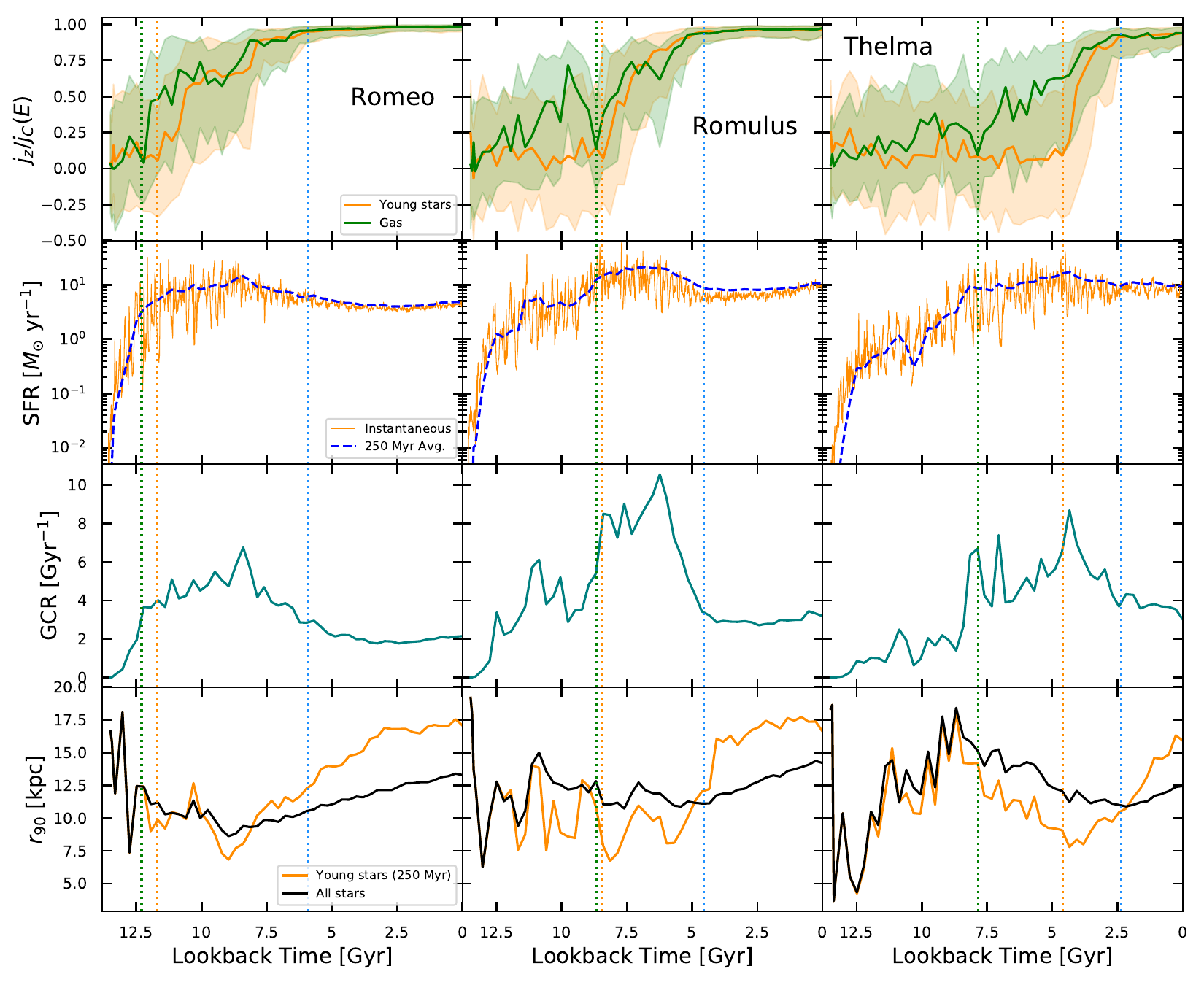}
    \caption{Relationship between kinematic phases, star formation, gas consumption, and the spatial growth of the stellar component. The top row shows the evolution in circularity for young stars (orange) and gas (green) as a function of lookback time for three example systems. The second row shows the star formation rate (SFR); instantaneous rates are shown in orange and 250~Myr-averaged rates in blue. The third row shows the cool gas consumption rate (GCR), defined as the 250~Myr-averaged SFR divided by the mass in cool gas ($T<10^4$~K) within 20~kpc. The bottom row shows the radius enclosing 90\% of the stellar mass within 20~kpc, $r_{90}$, for all stars (black) and young stars ($\leq 250$~Myr; orange). In every panel, vertical dotted lines mark gas spin-up (green), stellar spin-up (orange), and cooldown (blue). Star formation becomes steady after cooldown in each system, while the GCR peaks between spin-up and cooldown. The young-star $r_{90}$ is generally compact during the high-GCR phase and grows outward afterward, indicating that the recently star-forming component becomes increasingly extended as the system approaches the thin-disk phase. The peak of the GCR in each galaxy coincides with the time that $r_{90}$ starts growing monotonically up to the present day.}
\label{fig:starformation}
\end{figure*}

\begin{figure*}
    \centering 
    \includegraphics[width = \textwidth]{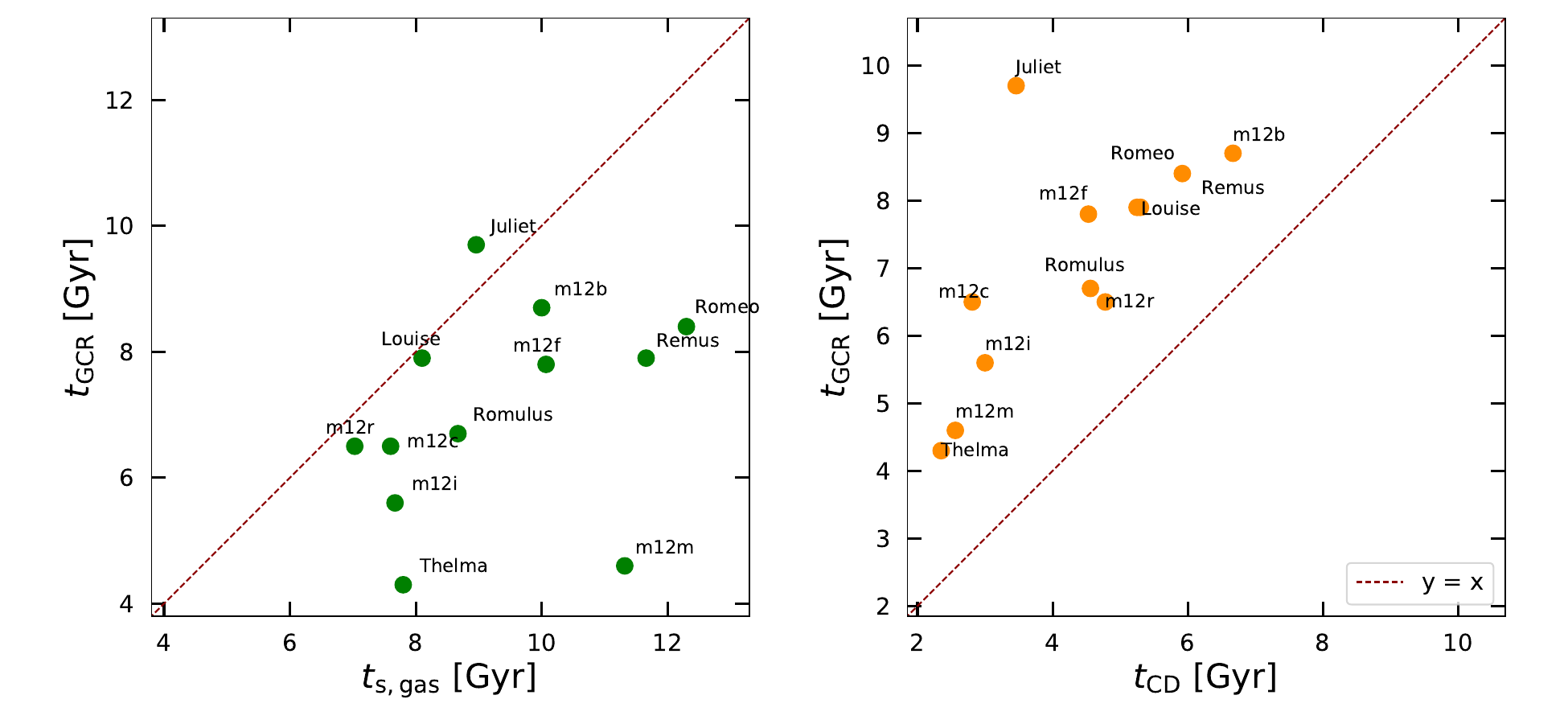}
    \caption{The Gas Consumption Rate (GCR) into stars peaks between spin-up and cooldown.  Left: time when the GCR peaks plotted versus the gas spin-up time for each galaxy.  Here GCR is defined as in Figure \ref{fig:starformation}. Right: The GCR peak time plotted versus the cooldown time for each galaxy.  In both panels, the red dotted lines show one-to-one relations.  In all cases except Juliet, the gas consumption rate peaks after gas spin-up and before cooldown. }
\label{fig:SFE scatter}
\end{figure*}

A more direct comparison of the different transition timescales is provided in Figure \ref{fig:t_gas t_star t_cool}.  The left panel presents a scatter plot of the (lookback) gas spin-up times versus stellar spin-up times for each galaxy.  The dashed line shows the one-to-one relationship. We see that the two timescales follow each other roughly linearly, with the gas spinning up $\sim 1$ Gyr earlier on average (with a longer lookback time to spin-up). The right panel plots stellar spin-up time versus cooldown time.  Here we see that cooldown is always later than stellar spin-up (with a smaller lookback time) and that the correlation between the two is quite weak. Observationally, this matches the expectation that the thick disk should be older than the thin disk.  The fact that the two times are otherwise uncorrelated suggests that the physics driving the two transitions is likely distinct.

\section{Empirical Trends}
\label{sec:empirical-trends}
\subsection{Mass Growth, Gas Fraction, and Mergers}

We are interested in understanding whether the spin-up and cooldown times identified above correlate with straightforward quantities such as virial mass, stellar mass, and galaxy gas fraction, and whether significant mergers (which create sharp jumps in these quantities) also play a role. Figure~\ref{fig:characteristic masses and gas fraction} explores just this. We show the evolution of each galaxy's virial mass (top), stellar mass (middle), and gas fraction ($f_{\rm gas} = M_{\rm gas} / (M_{\rm gas} + M_\bigstar)$ within 20 kpc, bottom). The color of each line maps to a specific simulation, as indicated in the legend.  Along each line, the circles, squares, and $\times$'s mark the associated times for gas spin-up, stellar spin-up, and cooldown, respectively.  The top panel shows the well-known result that galaxy halos tend to grow rapidly in virial mass at early times with slower growth at late times \citep{Wechsler02}. 

We see that galaxies undergo stellar spin-up at virial masses $\sim 3-10\times10^{11} M_\odot$ and stellar masses $\sim 3-10 \times 10^{9} M_\odot$.  There does not appear to be a sharp characteristic virial mass or stellar mass that coincides with the transitions.  These findings are consistent with the results of other simulation projects \citep[e.g.][]{artemis, TNG50disks}. 

A complementary way to view the stellar mass growth is to ask what fraction of each galaxy's present-day stellar mass is accumulated during the protogalaxy, thick disk, and thin disk phases. We present this decomposition in Appendix~\ref{sec:phase_masses}. The phase mass fractions vary substantially across the suite and show no clean monotonic trend with present-day stellar mass; instead, they primarily reflect the relative durations of the three kinematic epochs.

It is notable that the virial and stellar masses continue to increase after spin-up, but the growth rate is typically fairly gradual.  There are, however, a few instances where distinct mergers are evidenced by jumps in mass, seen first in the virial mass lines and later in stellar mass.  This delay corresponds to the time it takes for infalling galaxies to travel from the virial radius to the central galaxy. The clearest examples are Romulus, with an early merger, and m12r, with a late merger. Importantly, these mergers {\em do not} appear to correlate in any systematic way with any of the characteristic timescales of the kinematic transitions. This conclusion is also supported visually by Appendix \ref{sec: mergers}, which shows the gas and young-star morphologies of Romeo, Romulus, and Thelma near their respective gas and stellar spin-up times. In all three cases, the gas is irregular, filamentary, and clumpy, with visible accretion streams and disturbed structure. The young-star distributions, however, are already centrally concentrated and do not show an obvious merging stellar companion of comparable mass at the time of spin-up. Thus, while anisotropic gas accretion and minor interactions may contribute to the chaotic conditions from which the early disk emerges, these morphologies do not provide clear evidence that spin-up is triggered by a discrete major merger.

Beyond mergers, the three ELVIS Local Group analog pairs in our sample (Romeo/Juliet, Romulus/Remus, Thelma/Louise) provide a direct test of whether the kinematic transitions are driven by shared large-scale environment. Each pair separately occupies the same zoom-in volume and therefore experiences similar large-scale cosmological accretion conditions, yet the pair-mates do not consistently produce matched spin-up or cooldown histories. The contrast is most striking for Romeo and Juliet, whose stellar spin-up times differ by nearly 4 Gyr (11.7 versus 7.75 Gyr lookback) and whose protogalactic stellar-mass fractions differ by almost an order of magnitude (6\% versus 43\%; see Appendix~\ref{sec:phase_masses}). That galaxies embedded in the same large-scale environment produce such different kinematic histories supports the interpretation that each galaxy's internal history primarily governs its transitions between phases rather than larger-scale external environmental factors. This is consistent with observational evidence that the Milky Way and M31 have significantly different evolutionary histories despite a shared large-scale environment \citep{Yin2009}.

We define the inner gas fraction within 20 kpc as $f_{\rm gas} = M_{\rm gas} / (M_{\rm gas} + M_\bigstar)$, and plot its evolution in the bottom panel of Figure \ref{fig:characteristic masses and gas fraction}. For all simulated galaxies, 
$f_{\rm gas}$ follows a similar path in its evolution. Gas fractions drop rapidly from near unity and become strikingly constant at late times.  In all but one case, the gas fractions stabilize just after cooldown, when the star formation rate balances the gas accretion rate in the long term, because large-scale outflows are suppressed following cooldown, or the end of bursty star formation in the terminology of previous FIRE papers \citep{Stern_2021, Gurvich23}. The lone exception is m12r, which has a starburst just after a late-time merger, driving the gas fraction sharply downward in the last billion years of evolution. 

\begin{figure*}
    \centering
    \includegraphics[width = \textwidth]{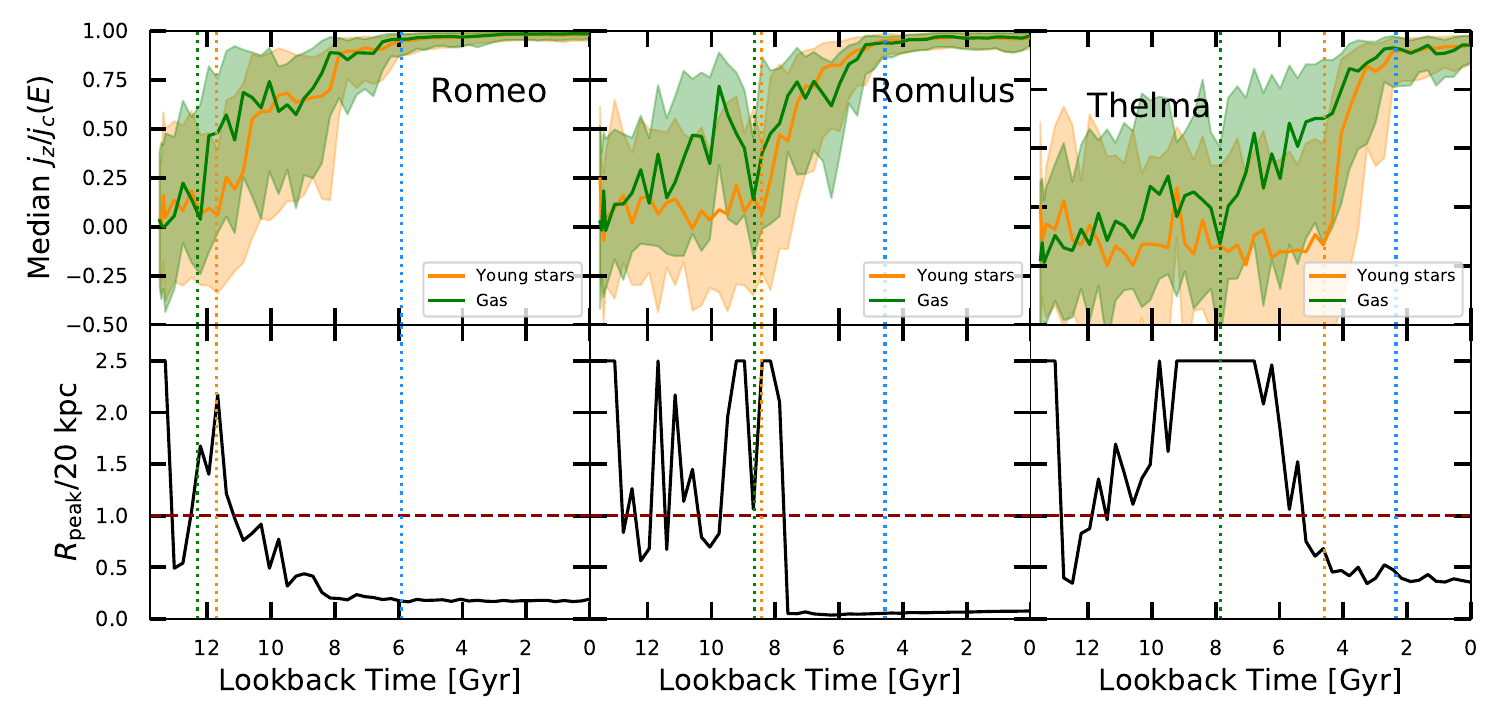}
    \caption{Does potential concentration drive spin-up? Shown is the time evolution of the circularity of young stars (top row), along with a measure of the potential concentration, $R_{\rm peak}/20$ kpc (bottom row), for three galaxies in our simulation suite that possess MW masses at present day. We label the spin-up times for these galaxies in gold and cooldown times in blue. For these selected galaxies, we observe that $R_{\rm peak}/20$ kpc reaches a local maximum in its time evolution very near the spin-up time and, importantly, has a value above unity, indicating a centrally \textit{un}concentrated potential when the disk starts forming.}
    \label{fig:jzjc_rpeak20}
\end{figure*}

\subsection{Star Formation and Galaxy Size}
\label{subsec:star-formation}
As discussed in the Introduction, past work with FIRE-2 simulations has seen links between the evolution of star formation burstiness over cosmic time and the transition from thick to thin disk formation \citep{Yu2021, Gurvich23}. This link between the nature of star formation and morphology may arise from an evolution in galactic dynamical timescales versus the timescale of stellar feedback \citep{CAFG2018}, inner CGM virialization \citep{Stern_2021}, and/or steepness of the central potential \citep{Hopkins_2023}.

Figure~\ref{fig:starformation} explores the connection between star formation and kinematic phases, now in the context of the spin-up and cooldown times. The top row shows evolution in the circularity of young stars (orange) and gas (green) as a function of lookback time for three example halos: Romeo, Romulus, and Thelma, from left to right. As in previous figures, solid lines show median circularity, and the shaded bands show the 90th-percentile range. The green, orange, and blue vertical dotted lines mark the gas spin-up, stellar spin-up, and cooldown times, respectively.

The second row in Figure~\ref{fig:starformation} shows the `instantaneous' star formation rate (SFR) measured over 10~Myr (orange) and averaged over 250~Myr (blue), each measured within a spherical region of 20~kpc from the host center. As expected from previous work, star formation remains bursty before and after the spin-up times, but becomes steady after cooldown, during the thin disk phase.

The third row plots a different star formation metric: the gas consumption rate (GCR). Here, we define the GCR as the SFR, averaged over 250~Myr, divided by the cool gas ($\leq 10^4$~K) mass within 20~kpc:
\begin{align}
    {\rm GCR} = \frac{{\rm SFR}_{\rm 250\,Myr} (t)}{M_{{\rm gas},\, T \leq 10^4\,{\rm K}} (t)}.
    \label{eq:gcr}
\end{align}

Note that the GCR peaks between the gas spin-up times and the cooldown times in each case, i.e. the ratio of SFR to cool-gas mass is highest prior to cooldown, while star formation is still bursty.

The bottom row of Figure \ref{fig:starformation} shows the radius enclosing 90\% of the stellar mass within 20~kpc, $r_{90}$, for all stars (black) and young stars ($\leq 250$~Myr; orange). The young-star $r_{90}$ traces the radial extent of the recently formed stellar component. In these examples, the galaxies are relatively compact during the high-GCR phase and grow outward afterward, suggesting that sustained radial growth of the star-forming component follows the period of most rapid cool-gas consumption. The growth is seen most cleanly among young stars, suggesting an ``inside-out'' growth of the thin disk (but not the other components). In the case of Thelma, the galaxy actually shrinks in radial size after spin-up and only begins to get larger around the time of cooldown. A contraction of the star-forming disk around the thick-to-thin transition has also been inferred for the Milky Way: \citet{funakoshi2025thicktothindisctransitiongas} find a dip in the stellar scale length between the thick and thin disk populations and interpret it as a significant shrinking of the star-forming gas disk, which they associate with the end of the starburst accompanying the Gaia-Sausage-Enceladus merger. Radial-size changes of this kind occur in our suite without a coincident major merger, which is consistent with our broader finding that mergers are not required to drive the kinematic transitions.

Figure \ref{fig:SFE scatter} shows that the trend for the GCR to peak between the gas spin-up times and the cooldown times generally holds for our full sample of simulations. Specifically, on the left we plot the time when GCR peaks for each galaxy versus the gas spin-up times. On the right, we show the GCR peak time versus the cooldown time for each galaxy. Dashed red lines in each panel show one-to-one lines. On the left, we see that in all but one case (Juliet), the gas spins up before the gas consumption rate peaks. On the right, we see that in all cases the GCR peaks prior to cooldown. While in the case of Juliet, the absolute peak in GCR occurs just prior to spin-up, the qualitative trend that the GCR remains high after spin-up and drops after cooldown remains true.

In order to better understand these trends, in Appendix \ref{sec:gce temps} we have looked at the GCR defined using {\em cold} gas: the star formation rate per unit mass in $<100$ K gas, rather than the rate per mass of $<10^4$ K gas used in Eq. \ref{eq:gcr}.  We find that the GCR defined using cold gas is much higher at all times. In addition, the GCR defined using cold gas does {\em not} clearly peak during the spin-up/thick disk phase, but rather remains fairly high from early in the protogalaxy phase through the spin-up phase.  What is common, however, is that both GCRs drop after cooldown.  In a relative sense, the decrease in GCR of cold gas is even more pronounced after cooldown than we see in cool gas.  This may be because cold, dense gas in a thin disk can experience strong shearing or stretching from spiral arms prior to collapsing into a self-gravitating star-forming region -- phenomena less prevalent during the protogalaxy and thick-disk stages. Further, the later transition from bursty to steady star formation at cooldown is expected to bring the galaxy into a regime where stellar feedback regulates the effective star formation efficiency per free-fall time to a lower value \citep{Hopkins2011, CAFG2013, Gurvich23}, thus lowering the GCR to moderate levels.  An interesting direction in the future will involve connecting these kinematic phases more directly to the nature of star formation in the ISM and testing these ideas.

\section{Exploring Causation}
\label{sec:causation}

\subsection{Concentration of Central Potential}
\label{sec: concentration}

\citet{Ceverino17} and \citet{Dekel2020} have shown that disks often form in simulations after compaction events steepen the galaxy potential; they have argued that disks are promoted in this circumstance because the timescale for inward mass transport decreases when the mass profile is steep. In related work, recent studies have suggested that the central concentration of the gravitational potential plays a central, causal role in driving disk formation. Using a suite of controlled numerical experiments on the FIRE dwarf galaxy m11a restarted at $z=1$, \citet{Hopkins_2023} varied physical parameters (e.g. cooling, stellar feedback, and modified potentials) to isolate the mechanisms that promote disk growth. They demonstrated that m11a transitions into an angular momentum-dominated, disky system specifically when its gravitational potential is artificially modified to be centrally concentrated (producing a peaky rotation curve). In contrast, when the rotation curve is rising (indicating an unconcentrated potential), the m11a system retains an isotropic distribution with little to no net angular momentum. \citet{Hopkins_2023} argued that this arises because a steep potential promotes a well-defined dynamical center, ceases to support the global `breathing modes' of feedback, and promotes orbit mixing with a stable disk. Based on these controlled experiments, they concluded that a centrally concentrated mass profile is a key trigger for the morphological transition into a disk.

These results suggest a natural hypothesis: does this physical trigger generalize to more massive systems in their natural cosmological evolution? In other words, might disk spin-up in Milky Way-mass systems also be triggered by the steepening of the central potential?

Figure \ref{fig:jzjc_rpeak20} presents an exploration into this question for our suite of cosmological FIRE-2 MW-mass (m12) galaxies. Shown along the top row is the circularity evolution of the same three galaxies presented in Figure \ref{fig:starformation}. To test the generalizability of the potential concentration trigger, the bottom row shows a measure of central concentration: the radius where the rotation curve peaks, $R_{\rm peak}$, in units of 20 kpc. If the peak radius is small compared to 20 kpc, then the potential is steep because the rotation curve is falling throughout most of the galaxy. If the peak radius is large compared to 20 kpc, that means the rotation curve is rising throughout the radii of interest and the potential is puffy. We plot this in units of 20 kpc since in this plot we are looking at the circularities of stars within 20 kpc. We find no qualitative difference if instead we examine the evolution of $R_{\rm peak}/R_{90}$, where $R_{90}$ is the 90th-percentile total mass radius (including stars, gas, and dark matter) of the galaxy.

We see from the bottom panel of Figure \ref{fig:jzjc_rpeak20} that the galaxy's transition to an angular momentum-dominated system consistently occurs when this ratio is still significantly above 1. In other words, our MW-mass galaxies begin forming an early disk when their potentials are still ``puffy'' and decidedly unconcentrated. In all three cases, the potential becomes steep ($R_{\rm peak}/ 20\,{\rm kpc}$ dips significantly below 1) only after gas spin-up and before disk cooldown (blue vertical lines). We find that this trend holds for our full suite. Consequently, while controlled restart experiments had previously suggested that potential concentration induces disk formation, we do not find evidence that it is the key trigger for the formation of galaxy disks in our sample of MW-mass FIRE galaxies evolving cosmologically. The steepening of the central potential is not the trigger for spin-up, but rather occurs after spin-up and prior to the onset of the thin-disk phase.

This interpretation is broadly consistent with results from other simulation suites, though with some important differences. Using $v_{\mathrm{c}}(R_{\mathrm{eff}})/v_{\mathrm{c}}(3R_{\mathrm{eff}})$ as a measure of central concentration, \citet{artemis} found that disks in the ARTEMIS simulations generally form when the mass distribution is not especially centrally concentrated, and that the central concentration increases after disk formation. \citet{vadim2} found in TNG50 that disk formation is correlated with the steepening of the central potential, but emphasized that this correlation does not by itself establish causation, since substantial steepening occurs not before but after the onset of disk formation. Taken together, these results suggest that central concentration is closely associated with disk growth and stabilization, but is not necessarily the initial trigger of disk spin-up in simulations. The agreement among FIRE-2, ARTEMIS, and TNG50 therefore points to a common time ordering in which disk spin-up precedes the central potential reaching its most concentrated state.

Our interpretation of this result is that baryonic contraction \citep{bc_ref,Gnedin04} in our galaxies occurs only after spin-up and begins in earnest only during the thick disk phase. The associated steepening of the potential may be a necessary condition for cooldown and the onset of the thin disk phase. For example, it is possible that potential steepening helps usher in the virialization of the inner CGM, which is sensitive to the depth of the potential via the virial temperature \citep{Stern_2021, Byrne2023, Hopkins_2023} and tends to coincide with the cooldown time and give rise to the ability for the angular momentum of gas to become coherently aligned prior to joining the disk \citep{Stern_2021,Hafen_2022,Gurvich23,bornthisway, Stern2024}.  We revisit the trend with CGM virialization in the next subsection.

As a supplement to this section, please also see Appendix \ref{sec: central concentration appendix} for a detailed exploration of the evolution of the shape of Romeo's rotation curve and how it relates to the evolution of its stellar circularity distribution over time.

A more recent model of disk formation involving a potential-steepening scenario has been put forward by \citet{semenov2026}, who develops an analytic model in which gas disk formation occurs when the circular velocity at the baryonic half-mass scale becomes comparable to the velocity dispersion of accreting gas. In the cold-mode regime, where $\sigma\sim V_C(R_{\rm vir})$, this corresponds approximately to $V_C(R_{50,b})/V_C(R_{\rm vir})\sim 1$. In Appendix~\ref{sec: central concentration appendix}, we explore this parameter for the six ELVIS galaxies in our FIRE-2 sample. We find that $V_C(R_{50,b})/V_C(R_{\rm vir})$ is typically either near unity around gas spin-up or at a minimum value, and then rises substantially during the subsequent thick-disk phase.

\subsection{Inner CGM Virialization}
\label{subsec:CGM}

\begin{figure*}
    \centering
    \includegraphics[width = \textwidth]{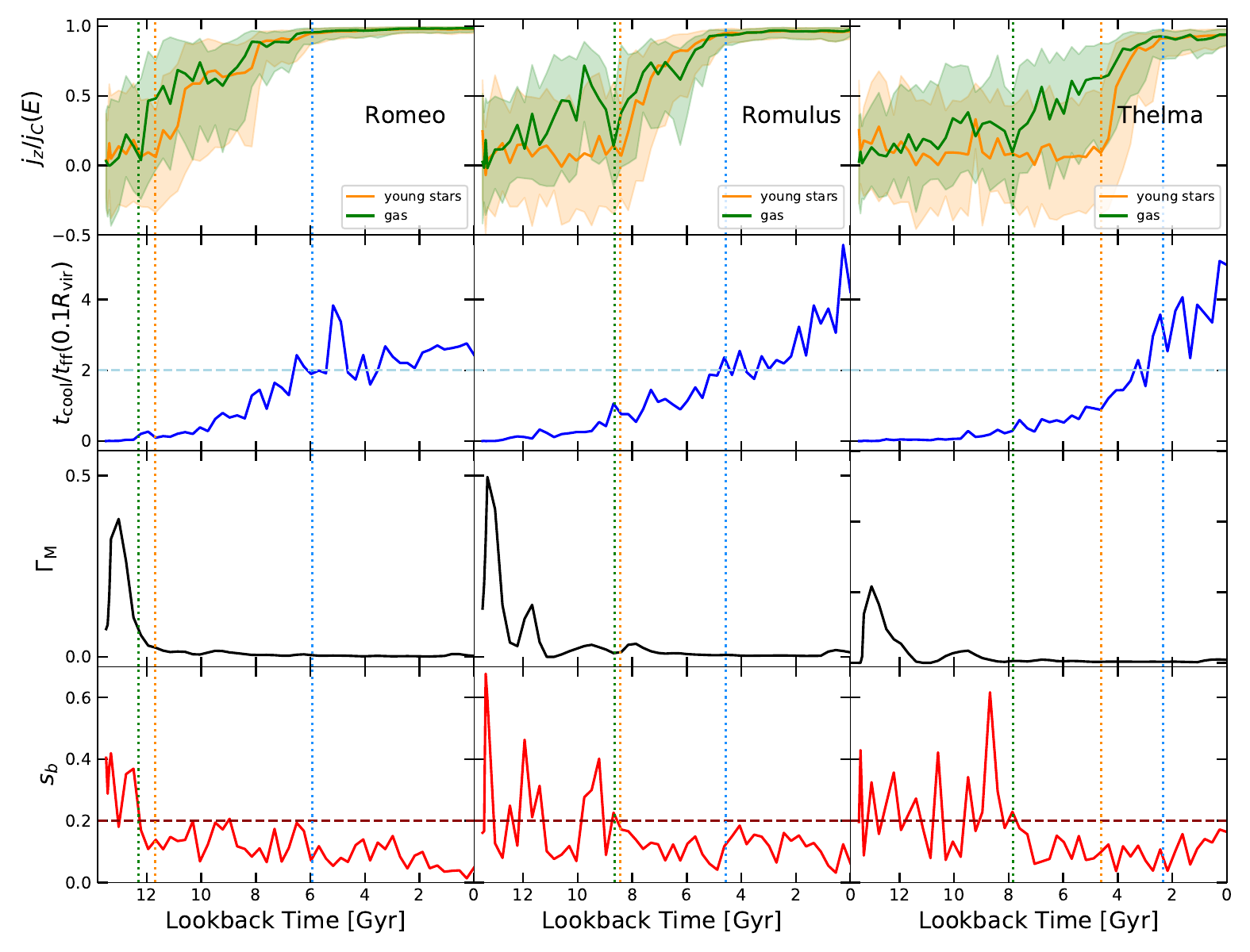}
    \caption{Inner CGM virialization, rapid accretion, and baryonic sloshing. The top row reproduces the circularity evolution of young stars (orange) and gas (green) shown in Figure \ref{fig:jzjc_rpeak20} for Romeo, Romulus, and Thelma.  The second row shows the ratio of the gas cooling time to the free-fall time in the inner galaxy.  This ratio characterizes the degree of virialization of the CGM, such that larger values of this ratio coincide with smoother and more thermally supported gas accreting onto the galaxy. We see that cooldown occurs at about the same time as the virialization of the inner CGM, when $t_{\rm cool}/t_{\rm ff}$ first rises above $\sim 2$. The third row shows a measure of the logarithmic mass growth rate within the central 20 kpc to the dynamical time within the same radius: $\Gamma_{\rm M}$  $\equiv  t_{\rm dyn} \times {\rm d}\ln M/{\rm d}t$. Large values of $\Gamma_{\rm M}$ mean that the mass grows significantly in a dynamical time. The bottom row shows the baryonic sloshing parameter $s_{\rm b}$, defined in the text, which quantifies the difference between baryonic motion and the total COM motion of the system. Gas spin-up occurs when $s_{\rm b} \lesssim 0.2$. We show this baryonic sloshing parameter for our full suite in Appendix \ref{sec:offset_full_suite}.}
    \label{fig:jzjc tdyn baryon}
\end{figure*}

As discussed above, previous work has shown that the onset of thin disk formation (cooldown) coincides with the virialization of the inner CGM \citep{Yu2021,Stern_2021,Hafen_2022,Gurvich23,bornthisway, Sultan2026}. 

For the sake of reproducibility and completeness, the top two rows of Figure \ref{fig:jzjc tdyn baryon} show this result.   The top row shows the circularity evolution for the three example galaxies shown in Figure \ref{fig:jzjc_rpeak20}. The vertical blue dotted line shows the cooldown time defined above.  The second row  plots a quantity that measures CGM virialization as the ratio of the cooling time of shocked gas $\tcoolsh$ to the free-fall time $\tff$ at an inner radius $r = 0.1~R_{\rm vir}$. This parameter was introduced by \cite{Stern_2021}, who showed that when $t_{\rm cool} /  t_{\rm ff}$ exceeds $\approx2$ the inner CGM ``virializes'', i.e. the inner halo becomes filled with a hot, thermal pressure-supported and quasi-static medium\footnote{See Stern et al. (2021) for a detailed discussion of how we evaluate the ratio $\tcoolsh / \tff$ in our simulations. In short, the cooling time in this ratio is a proxy for the expected cooling time of a hot virialized phase, which is not present before the CGM actually virializes, so the exact definition can be important to reproduce our results.}. Prior to this transition, the inner CGM is cool and supported mainly by turbulence \citep{kakoly2025}. \cite{bornthisway} showed that this virialization of the inner CGM coincides with the transition from bursty to steady galaxy star formation. Figure~\ref{fig:jzjc tdyn baryon} shows that the cooldown time also coincides with the time when $t_{\rm cool} / t_{\rm ff}$ first crosses 2, i.e. with inner CGM virialization, as found also by previous studies \citep{Stern_2021,Yu2021,Gurvich23,bornthisway}. The larger the ratio of $\tcoolsh / \tff$, the smoother the inner CGM is and the more strongly it is supported by thermal pressure. 

\subsection{Rapid Accretion, Centering, and Baryonic Sloshing}
\label{subsec:sloshing}

One point emphasized by \citet{Hopkins_2023} is that a well-defined dynamical center could play a role in facilitating disk spin-up. In their simulations, externally imposed mass concentrations at galactic centers tended to promote disk formation. In addition to steepening the central potential, adding mass at galaxy centers also creates a clear dynamical center for gas and stars to orbit around.

At early times, we expect galaxies and dark matter halos to accrete matter rapidly onto their central regions \citep[e.g.][]{Wechsler02}; it may be that if galaxies are growing fast compared to the local dynamical time, it is difficult for the system to reach the dynamical equilibrium necessary for coherent rotation.

The third row of Figure \ref{fig:jzjc tdyn baryon} shows the relative mass accretion rate. Specifically, we define the relative mass accretion rate as 
\begin{equation}
\Gamma_{\rm M} \equiv   \frac{{\rm d}\ln M}{{\rm d}t} \times t_{\rm dyn},
\end{equation}
where $M$ is the total mass within 20 kpc and $t_{\rm dyn}$ is the dynamical time inferred from the total mass within 20 kpc. Large values of $\Gamma_{\rm M}$ mean that the mass grows significantly in a dynamical time.  In all three cases, the relative growth rate is high prior to spin-up, though in some cases the time delay between peak relative growth rate and spin-up is more than 2 Gyr. This suggests that the cessation of rapid mass accretion is not a discrete trigger for spin-up.

A related halo-assembly diagnostic was used by \citet{Ma2026}, who define the specific growth rate of the virial velocity as $\gamma \equiv \mathrm{d}\ln V_{\rm vir}/\mathrm{d}\ln a$. They identify a fast-to-slow halo transition bracketed by $\gamma = 3/8$ and $\gamma = 0$, and show that this transition correlates with changes in thin-disk fraction, dynamical hotness, and star-formation burstiness in FIRE-2 Milky Way-mass galaxies. Our $\Gamma_{\rm M}$ differs in that it measures the growth of the total mass within the inner 20 kpc, normalized by the local dynamical time, rather than the growth of the host halo virial velocity. Nevertheless, both diagnostics test the same broad idea: whether rapid mass assembly prevents the galaxy from reaching the dynamical equilibrium needed for coherent disk formation.

Rapid mass accretion is not the only possible source of angular momentum perturbation.  Anisotropic cosmological stream accretion, asymmetric feedback, massive clumps, or mergers could all deliver momentum impulses on timescales that perturb an orbit. If we consider gas attempting to orbit about a center whose bulk velocity differs from the bulk velocity of the baryonic material itself by a value $\Delta V$, then the expected angular momentum perturbation at a galactic radius $R$ associated with such sloshing is $\delta j \sim R \Delta V$.  If we compare this to a typical value of specific angular momentum at the same radius, $j \sim R V_c(R)$, we can write a dimensionless parameter  $s_{\rm b} \sim \delta j/j \sim \Delta V/V_c(R)$.

Concretely, we define a ``sloshing'' metric that  measures the offset between the total center-of-mass motion of the system and the motion of baryons:
\begin{align}
    s_{\rm b} = \frac{|\vec{V}_{\rm baryons} - \vec{V}_{\rm total}|}{V_{c} (R_{90})},
    \label{baryon sloshing}
\end{align}
where $\vec{V}_{\rm baryons}$ is the mass-weighted 3D velocity vector of all of the baryonic matter within 20 kpc of the galaxy, and $\vec{V}_{\rm total}$ is the total velocity vector of all the mass, including dark matter, within the same region. The difference between the baryonic velocity and total velocity is then normalized by the circular velocity, $V_{c} = \sqrt{G M/R}$ at $R=R_{90}$ (the radius that encloses 90\% of the stellar mass within 20 kpc).


The bottom panel of Figure \ref{fig:jzjc tdyn baryon} plots $s_{\rm b}$ as a function of time for each galaxy. As might be expected, the sloshing variable is higher at early times and begins to drop towards lower values at late times.  In each case,  $s_{\rm b}$ drops below a value of $0.2$ at approximately the time of gas spin-up, which suggests that having a stable center may be crucial for enabling coherent spin. 

Based on this observation, we define a ``sloshing lookback time'', $t_{\rm slosh}$, to be the time when the dimensionless COM motion offset falls below $0.2$ of the circular velocity.  Figure \ref{fig:t_slosh_t_s,gas} shows this time plotted versus the spin-up time for gas in each galaxy. In all but one case (Remus), the relative COM motion settles down prior to gas spin-up.  This suggests a scenario where, at early times, the system is too dynamically chaotic to support coherent rotation. We plot $s_{\rm b}$ and $j_z/j_C(E)$ for the full simulation suite in Appendix \ref{sec:offset_full_suite}, showing that this COM motion is erratic and chaotic before gas spins up in each galaxy.

What, then, causes $s_{\rm b}$ to decline? Our diagnostics do not identify a single cause, but they suggest that the decline reflects a gradual reduction in the large-scale disturbances that move the baryons coherently relative to the total mass distribution. This decline could be driven by starbursts that drive gas coherently in a single direction, or mergers that deliver momentum over a timescale that is similar to an orbital time. All of these processes are likely important during the protogalaxy phase, when mass assembly is rapid, star formation is bursty, and mergers and strongly anisotropic accretion are more common. Determining the relative contributions of accretion, feedback, and mergers will require a more direct decomposition of the associated mass and momentum exchange.

\begin{figure}
    \centering 
    \includegraphics[width=1 \columnwidth]{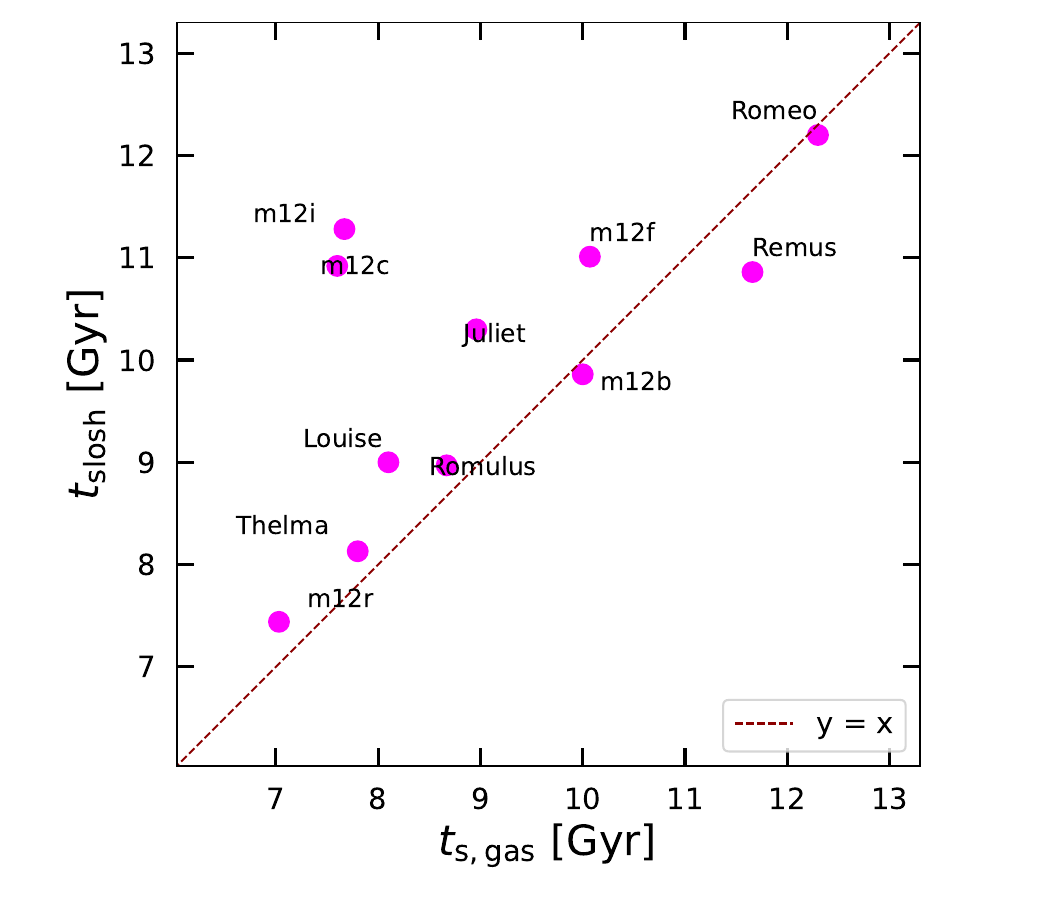}
    \caption[]{The ``sloshing time'' -- defined as the lookback time when the offset between the COM motion of the baryons and the total mass is smaller than 20\% of the circular velocity -- plotted versus the gas spin-up time for each galaxy.  We see that in almost all cases, the center-of-mass motion settles down prior to gas spin-up.}
    \label{fig:t_slosh_t_s,gas}
\end{figure}

\subsection{Distinct Physics of Spin-up and Cooldown}
\label{subsec:synthesis}

Taken together, the results in Sections~\ref{sec: concentration}--\ref{subsec:CGM}, together with the sloshing analysis above, suggest that spin-up and cooldown are not two instances of the same physical transition. Instead, they appear to mark two different stages in the dynamical settling of MW-mass galaxies. The first transition, from the protogalaxy to the thick disk, occurs once the galaxy develops a sufficiently stable dynamical center for gas to orbit coherently. The second transition, from the thick disk to the thin disk, occurs only later, after the inner CGM and central potential have evolved into a regime that allows accreting gas to join the disk with a more coherently aligned direction of angular momentum.

The onset of gas spin-up is most closely associated with the decline of baryonic sloshing. In 11 of 12 galaxies, the baryonic sloshing parameter $s_{\rm b}$ falls below $\sim 0.2$ at or shortly before gas spin-up (Figures~\ref{fig:jzjc tdyn baryon} and \ref{fig:t_slosh_t_s,gas}). By contrast, several other quantities that might plausibly control disk formation do not appear to reach their late-time configurations at spin-up. The central potentials are still relatively unconcentrated, with rising rotation curves (Figure~\ref{fig:jzjc_rpeak20}, see also Appendix~\ref{sec: central concentration appendix}), and the inner CGM has not yet virialized, with $t_{\rm cool}/t_{\rm ff}$ remaining below the threshold value of $\sim 2$. Coherent gas rotation can therefore emerge while the galaxy is still rapidly evolving, cool-mode accreting, and dynamically far from its late-time thin-disk state. A similar argument applies to clumpiness and feedback: star formation remains bursty and the emerging disk remains clumpy after spin-up \citep{bornthisway}; therefore, although clumpy star formation may contribute to early baryonic disequilibrium, neither the disappearance of clumps nor the end of bursty feedback can be what permits coherent gas rotation to emerge. \par

Cooldown appears to require a more restrictive set of conditions. The thick-to-thin disk transition coincides with several linked transitions: star formation becomes steady, the inner CGM virializes as $t_{\rm cool}/t_{\rm ff}$ crosses $\sim 2$ (discussed in Section~\ref{subsec:CGM} and shown in Figure~\ref{fig:jzjc tdyn baryon}), the central potential steepens as the baryonic mass distribution becomes more concentrated (discussed in Section~\ref{sec: concentration} and shown in Figure~\ref{fig:jzjc_rpeak20}), and accreting gas is expected to become more angular-momentum coherent before reaching the disk \citep{Stern_2021, Hafen_2022}. Our results indicate that thin disk formation does not occur in our suite until these conditions are broadly in place, whereas thick disk formation begins before they are established.


This difference in conditions has a simple physical interpretation. A thick disk requires coherent net rotation, but not a dynamically cold configuration. Once the baryons have a stable center of mass motion, gas can begin to rotate coherently even if gas continues to arrive on a broad range of orbits, naturally producing the wide circularity distributions characteristic of the thick-disk phase. 
A thin disk is more demanding: the gas that forms new stars must not only rotate, but must join the existing disk with a relatively well-aligned angular momentum direction and low orbital dispersion.  Such coherence is expected only once accretion through the inner CGM becomes slow enough, and the CGM sufficiently hot and pressure-supported, for gas to settle before reaching the star-forming disk \citep{Hafen_2022}. 
We can understand this by comparing the mixing time for angular momentum to become coherent, $t_{\rm mix} \sim r/c_s \sim r/V_c \sim t_{\rm ff}$, to the infall time, $t_{\rm infall}$. Importantly, when the CGM is hot, $t_{\rm infall} \sim t_{\rm cool}$.  This means that the virialization parameter is equivalent to the number of mixing times:  $N_{\rm mix} = t_{\rm infall}/t_{\rm mix}= t_{\rm cool}/t_{\rm ff}.$ That is, if the inner CGM is virialized, then the gas is expected to join the disk with coherent angular momentum. Prior to that time, when $N_{\rm mix} \lesssim 1$, it does not join the disk with coherent spin. 

 In this sense, spin-up is primarily associated with the emergence of a stable dynamical center within the galaxy, whereas cooldown is associated with the thermodynamic and angular-momentum structure of the gas before it joins the galaxy.

\section{Conclusions}
\label{sec:conclusion}

We have used a FIRE-2 simulation suite of 12 MW-mass galaxies that form disk galaxies at $z=0$ to explore their kinematic phases of evolution.   We build upon existing literature that has shown that the Milky Way appears to have evolved through three kinematic phases: a disordered protogalaxy, a coherently rotating thick disk, and a thin disk \citep{aurora,Chandra2024}.    In particular, \citet{Chandra2024} used the distribution of orbital circularities ($\epsilon$) of stars at fixed metallicity to identify these phases, and assumed metallicity was a proxy for stellar age to interpret the results. 
Figure \ref{fig:gaia and romeo three phases} shows that the same phases stand out in our simulations using the same approach. 
Figure \ref{fig:jzjc pdf scatter} shows a monotonic median age–metallicity relation in Romeo, which suggests that the observational approach of using [Fe/H] as a rough proxy for stellar age is reasonable.

With this preliminary analysis as motivation, we go on to study the time-sequence evolution of our galaxies, using the circularity distributions of both gas and young stars over time to identify three characteristic times that occur as a sequence in each galaxy: 1) the gas ``spin-up'' time, 2) the stellar ``spin-up'' time, and 3) the ``cooldown'' time (see Figures \ref{fig:romeo median and 1sigma}, \ref{fig:fire suite jzjc medians}, and \ref{fig:t_gas t_star t_cool}). Prior to the gas spin-up time, neither the stars nor cool gas have any coherent rotation ($\epsilon \sim 0$).  After gas spin-up, the median circularity of the gas begins to rise steadily, and after stellar spin-up the median circularity of the young stars begins to rise steadily.  Importantly, prior to the cooldown time, the spread in circularities remains broad, indicating that the orbits are not well aligned in a thin disk, but rather portray thick disk-like kinematics.  After cooldown, both stars and gas have $\epsilon \sim 1$ with a narrow spread.  This is indicative of thin disk orbits. Note that the cooldown time does {\em not} refer to the time when a pre-existing thick stellar disk becomes dynamically thinner.  Instead, it refers to a time after which {\em new} stars form in a thin disk.  Any pre-existing stellar thick disk remains thick.

In addition, we find that the nature of star formation correlates with kinematic phase. In previous work, \citet{Yu2021} showed that star formation tends to be bursty during the thick disk phase and transitions abruptly to steady star formation at the onset of the thin disk phase. To restate in the language of this paper: previous work has shown that star formation is bursty prior to cooldown and steady after cooldown \citep{bornthisway, Gurvich23, Stern_2021}.  In this paper we found additional trends related to the gas consumption rate and kinematic phase. 
In all of our galaxies,  the overall rate of consumption of cool gas converting into stars is {\em low} and the star-formation rate is {\em bursty} during the early disordered phase (prior to spin-up, see Figure \ref{fig:starformation}). After spin-up but before cooldown (during the thick disk phase), the rate of consumption of cool gas converting into stars is {\em highest} while the star formation rate remains {\em bursty} (see Figure \ref{fig:SFE scatter}).  After cooldown, during the thin disk phase, the star formation rate becomes {\em steady} and the cool gas consumption rate is intermediate (see Figure \ref{fig:starformation}). The thin disk phase also marks a time when inside-out growth of the disk emerges; prior to that time, young stars tend to have the same radii as older populations (see bottom row of Figure \ref{fig:starformation}). As discussed in Section~\ref{subsec:star-formation}, this evolution is plausibly a byproduct of the geometric transition at spin-up and the emergence of feedback self-regulation at cooldown, rather than a driver of either transition. Nevertheless, the link between kinematic phases, star formation rates,  gas consumption rates, and galaxy sizes merits future exploration.

One question we explored in some detail is why galaxies spin up in the first place.  As shown in Figure  \ref{fig:characteristic masses and gas fraction}, spin-up in our galaxies does {\em not} coincide with a specific gas fraction (ranges from $f_{\rm gas} \sim 0.3 - 0.9)$, nor a specific stellar mass (ranges from $M_\star \sim 3-10\times 10^{9} M_\odot$), nor a specific virial mass (ranges from $M_{\rm vir} \sim 3-10\times 10^{11}\,M_\odot$). Similarly, mergers do not appear to be a systematic trigger. We also looked to see if the central potential became highly concentrated prior to spin-up (see Figure \ref{fig:jzjc_rpeak20}), and instead found that galaxy potentials tend to be least concentrated, with rising rotation curves, at the time of spin-up. 

Instead, the spin-up times in our runs are better correlated with the measured difference between center of mass (COM) motion of the baryons and COM motion of the total potential. We refer to such an offset velocity $\Delta V$ as ``sloshing,'' which could be driven by asymmetric accretion or feedback. Specifically, for material at radius $R$ attempting to orbit about the dynamical center, a sloshing velocity acts as a perturbation in angular momentum: $\delta j \sim R \, \Delta V$. 
We find that the gas in the galaxy begins to spin coherently just after the sloshing phase ends (when $s_{\rm b} = \Delta V/V_c \sim \delta j/j \lesssim 0.2$), followed by spin-up of young stars (see Figures \ref{fig:jzjc tdyn baryon} and \ref{fig:t_slosh_t_s,gas}). 
  This suggests that a well-defined dynamical center for both the gas and total mass may be a prerequisite for the spin-up phase that prompts (thick) disk formation.

A complementary perspective on this process is provided by \citet{Bland-Hawthorn2025}, who study turbulent, gas-rich disks meant to resemble high-redshift Milky Way progenitors. They find that clustered star-formation feedback can drive a stochastic, Brownian-like displacement of the baryonic potential minimum relative to the total gravitational potential, with larger sloshing amplitudes in more gas-rich disks.
Their diagnostic is not identical to ours: they measure a spatial offset between potential minima, whereas $s_{\rm b}$ measures a velocity offset between the baryonic and total center-of-mass motion. Nevertheless, both quantities probe the degree to which the baryons fail to move coherently with the full mass distribution. Their results therefore support the interpretation that feedback or accretion-driven disequilibrium in gas-rich systems can contribute to the sloshing behavior captured by $s_{\rm b}$.

The sloshing we identify here is likely connected to the complicated  gas morphologies we see in these early systems.  As shown in Appendix \ref{sec: mergers}, they are irregular, filamentary, and clumpy near spin-up.  In a separate massive, gas-rich high-redshift FIRE galaxy, \citet{Oklopcic2017} found that individual massive clumps were short-lived (clumps above $10^8\,{\rm M_\odot}$ had a mean lifetime of $\sim 22$ Myr) and were generally disrupted by their own stellar feedback, even though the overall clumpy phase persisted for more than 1 Gyr. The repeated formation and feedback-driven disruption of such structures could redistribute baryonic mass and momentum asymmetrically, contributing to an asymmetric baryonic potential and to the behavior captured by $s_{\rm b}$. We do not explicitly identify or track individual clumps here, however, so their contribution cannot be separated from those of rapid accretion, mergers, and other sources of baryon disequilibrium.

The transition from the thick disk phase to the thin disk phase at cooldown has been the topic of previous work.  As mentioned in the introduction, \citet{Stern_2021} and subsequently \citet{bornthisway} showed that the transition to the thin disk phase coincides with the virialization of the inner circumgalactic medium (CGM, see also Figure \ref{fig:jzjc tdyn baryon}).   \citet{Hafen_2022} showed that once the inner CGM becomes virialized,  the angular momentum of infalling material becomes highly aligned prior to joining the galaxy. More recently, \cite{Sultan2026} confirmed this result in a broader suite of simulations including more massive halos whose inner CGM virializes at high redshift. Interestingly, we found that the central potentials of our galaxies become most concentrated prior to cooldown (see Figure \ref{fig:jzjc_rpeak20}).  Such a steepening of the central mass concentration could contribute to the onset of inner CGM virialization, perhaps via increasing the temperature and cooling time of shocked hot gas. 

There is an extensive body of literature on the role that mergers, especially gas-rich mergers, can play in disk formation \citep[e.g.][]{2006ApJ...645..986R, Brook2007,Kannan15, 2017A&A...600A..25R,  2020MNRAS.493.1375P}. We find no evidence in the mass accretion histories of our galaxies that suggests that distinct mergers systematically lead to spin-up. Prior to spin-up, our galaxies are growing rapidly in mass, and mergers are fairly common. In most cases, it is only after this period of rapid mass growth ends that spin-up occurs.  One exception is in the galaxy Romulus, which does have a significant influx of mass and gas, consistent with a gas-rich merger,  just prior to spin-up (see Figure \ref{fig:characteristic masses and gas fraction}).  Nevertheless, mergers of this kind do not systematically correlate with galaxy spin-up across our suite of simulations.

A separate question is whether mergers can contribute to the cooldown transition itself. \cite{Chandra2024} describe a different pathway in a TNG50 Milky Way analog, in which a major gas-rich merger at a lookback time of $\approx 7$ Gyr heats, thickens, and reorients a pre-existing high-$\alpha$ disk, while the subsequently accreted gas contributes substantial angular momentum and helps set the orientation of the present-day cold disk. In that case, the merger is not associated with the initial spin-up of the galaxy, but it is plausibly connected to the later thick-to-thin transition. We do not find clear evidence for an analogous merger-driven cooldown in our FIRE-2 sample. A distinct cooldown occurs in all 12 galaxies, and in most cases there is no obvious coincident major gas-rich merger in the mass accretion histories of Figure \ref{fig:characteristic masses and gas fraction}. The clearest counterexample is m12r, which undergoes a major merger only $\sim 1$ Gyr before $z=0$, several Gyr after its thin disk had already formed. In this case, the merger drives the median circularity downward rather than triggering the onset of thin-disk kinematics. 
%
%
The ubiquity of cooldown across our suite suggests that gas-rich mergers of the kind highlighted by \cite{Chandra2024} are not a necessary condition for three-phase kinematic evolution.

One result of potential observational interest is that coherent cool-gas rotation typically emerges before the spin-up of young stars (see Figures~\ref{fig:fire suite jzjc medians} and~\ref{fig:t_gas t_star t_cool}). In our suite, the offset between gas and stellar spin-up ranges from a few hundred Myr to more than 2 Gyr. 

This implies that Milky Way-mass progenitors generically pass through a phase in which rotationally supported cool gas coexists with kinematically broad or only weakly rotating young stars. With sufficient spatial and spectral resolution, this regime would appear as a galaxy with a coherent cool-gas velocity gradient but a dynamically hotter young-stellar component, distinct from both a fully disordered protogalaxy and a settled thin disk. Appendix~\ref{sec:velmap} shows this behavior explicitly for Thelma: after gas spin-up but before stellar spin-up, the cool gas has developed an ordered velocity gradient while the young-star velocity field remains incoherent. Recent JWST kinematic surveys at $z \gtrsim 2$ are beginning to probe this regime in cool gas \citep[e.g.,][]{Rowland24, Danhaive25}; analogous measurements of young-star kinematics in the same systems would test the predicted gas--star offset directly. More broadly, our results suggest that rotating gas alone is not sufficient evidence that a galaxy has reached the thin-disk phase: the key distinction is whether newly formed stars have also entered a dynamically cold, coherently rotating configuration.
This ordering is consistent with the results of \citet{TNG50disks}, who find that gaseous components are more rotationally supported than their stellar counterparts at fixed mass and redshift: gas disks form first, and the stellar disk follows.

This paper has presented an investigation into the physical origin of the key phases of disk evolution for Milky Way-mass galaxies.   In principle, a deeper understanding of the drivers of disk formation through simulations such as these can provide insights into tensions that may exist related to the formation of low-mass disks at $z=0$ in $\Lambda$CDM simulations \citep[e.g.][]{Klein25} and 
related tensions associated with the presence of early disks at  $z>2$   \citep[e.g.,][]{Ferreira2022,Robertson2023,Kartaltepe2023}. One key distinction we have made is that the onset of {\em thin disk} formation has a trigger that is distinct from the onset of {\em thick disk} formation. Thick disk spin-up appears to require primarily a coherent center of motion for gas to orbit around, while thin disk formation requires more: sub-sonic gas accretion enabled by inner CGM virialization and perhaps a steepening of the central potential. This suggests the need to sharpen the nature of existing tensions: do theoretical models need to produce more disks of any kind, or do they need more {\em thin} disks?  If it is the latter, for example, then processes that increase cooling times and/or enable more sub-sonic accretion would be required. 

It is likely that models need to be improved in order to reproduce the observed thinness of disks at early times, with recent work from JWST suggesting that thin disks may already be present at $z=3$ \citep{vanAsselt2026,Yu2026}. If so, conditions that allow for slower infall and more gas mixing prior to joining the galaxy may be needed. Longer cooling times or non-thermal support via cosmic ray heating could help.

 \par

\section*{Acknowledgements}
We thank Avishai Dekel and Joel Primack for many useful discussions that inspired our thinking on this work. They sadly passed away before we could share the final draft of this manuscript, but their imprint on how we and the community think about these topics will last well beyond the time we shared together. OM and JSB acknowledge support from NSF grant AST-2408246.  MBK acknowledges support from NSF grants AST-2108962 and AST-2408247; NASA grant 80NSSC22K0827; HST-GO-16686, HST-AR-17028, JWST-GO-03788, and JWST-AR-06278 from the Space Telescope Science Institute, which is operated by AURA, Inc., under NASA contract NAS5-26555; and from the Samuel T. and Fern Yanagisawa Regents Professorship in Astronomy at UT Austin. AW received support from NSF via CAREER award AST-2045928. JM is funded by a Pomona College Large Research Grant. FJM is funded by the National Science Foundation (NSF) Math and Physical Sciences (MPS) Award AST-2316748. CAFG was supported by NSF through grants AST-2108230 and AST-2307327; by NASA through grants 80NSSC22K0809, 80NSSC22K1124 and 80NSSC24K1224; by STScI through grant JWST-AR-03252.001-A; and by BSF through grant \#2024262.

\section*{Data Availability}

The data supporting the plots within this article are available on reasonable request to the corresponding author. A public version of the GIZMO code is available at \url{http://www.tapir.caltech.edu/~phopkins/Site/GIZMO.html}. FIRE data releases are publicly available at \url{http://flathub.flatironinstitute.org/fire} \citep{Wetzel2023, Wetzel2025}.



\bibliographystyle{mnras}
\bibliography{ref} 


\appendix

\section{Stellar Mass Growth Across the Three Kinematic Phases}
\label{sec:phase_masses}

\begin{figure}
    \centering
    \includegraphics[width=\columnwidth]{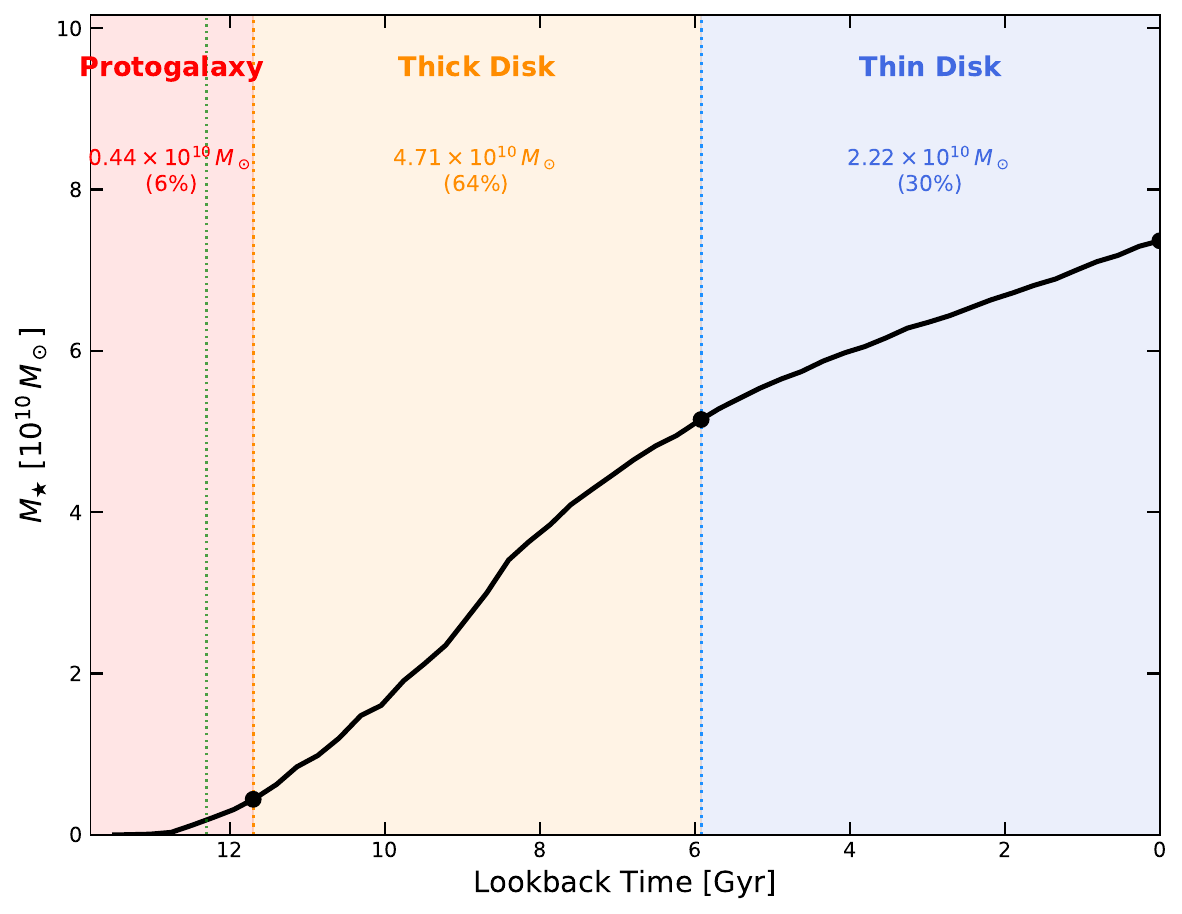}
    \caption{Cumulative stellar mass within 20 kpc of Romeo as a function of lookback time. Shaded regions mark the protogalaxy, thick disk, and thin disk phases, defined here using the stellar spin-up time and cooldown time. Vertical dotted lines mark the gas spin-up time (green), stellar spin-up time (orange), and cooldown time (blue). The gas spin-up time is shown for reference but is not used as a boundary in the stellar mass decomposition. Annotations give the net stellar mass growth within 20 kpc during each phase and the corresponding fraction of Romeo's present-day stellar mass within 20 kpc.}
    \label{fig:romeo_mass_phases}
\end{figure}

\begin{figure*}
    \centering
    \includegraphics[width=\textwidth]{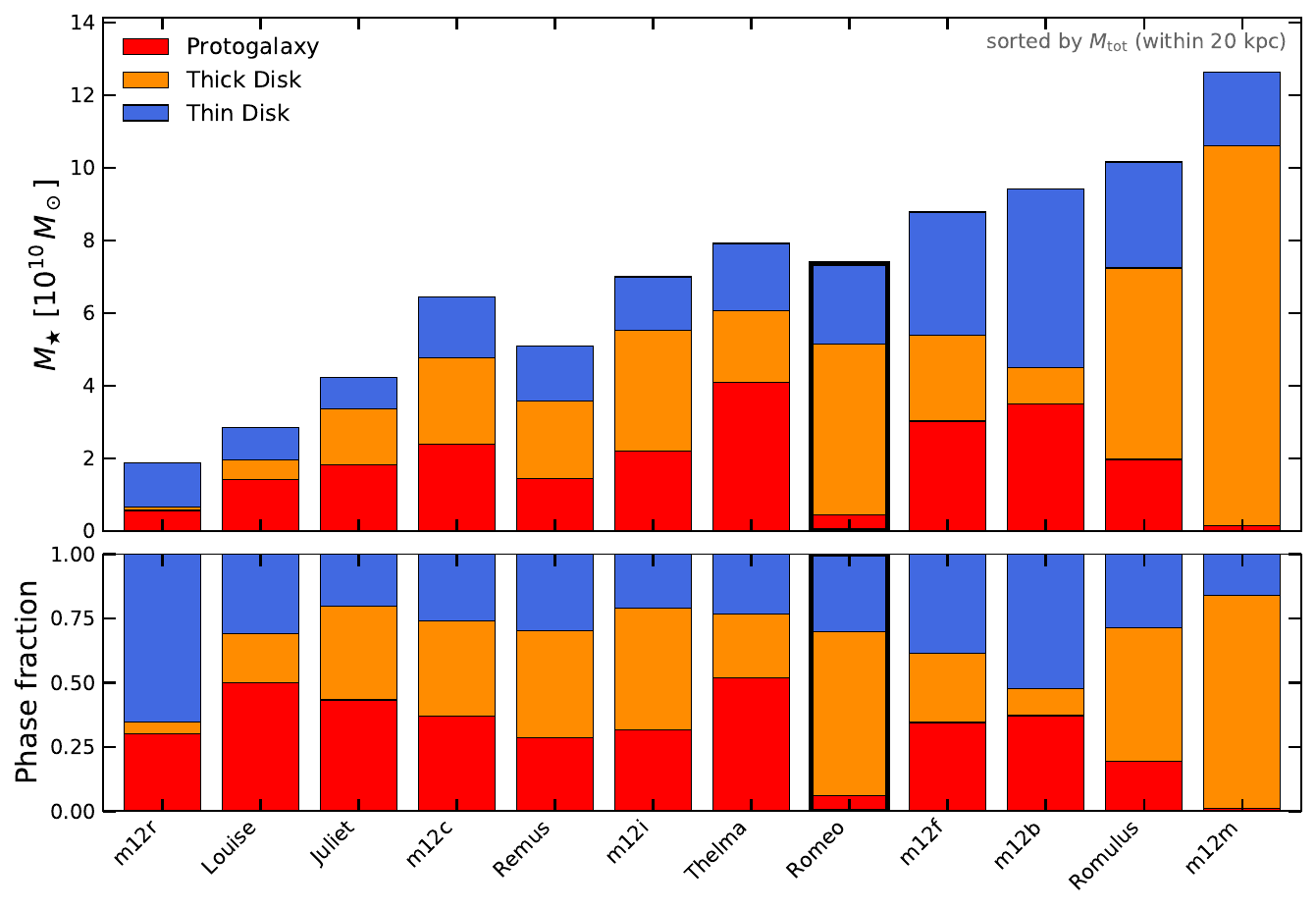}
    \caption{Stellar mass growth within 20 kpc across the three kinematic phases for the full simulation suite, sorted by total mass within 20 kpc. \textit{Top:} absolute stellar mass growth in each phase, shown as stacked bars. \textit{Bottom:} the same quantities normalized by each galaxy's present-day stellar mass within 20 kpc. Romeo, our prototypical example galaxy and MW-mass analogue,  is highlighted with a thick black outline. The relative mass growth associated with each phase varies substantially across the suite and shows no clean monotonic trend with total mass.}
    \label{fig:suite_mass_phases}
\end{figure*}

In Section~\ref{sec:sample-results}, we identified three kinematic phases in each galaxy: a disordered protogalaxy, a thick disk phase, and a thin disk phase. These phases are separated by the stellar spin-up time, $t_{\rm spin\textnormal{-}up,\star}$, and the cooldown time, $t_{\rm CD}$. The gas spin-up time, $t_{\rm spin\textnormal{-}up,gas}$, typically precedes stellar spin-up and is shown in Figure~\ref{fig:romeo_mass_phases} for context, but it is not used as a boundary in the stellar mass accounting below. A natural question is how much of each galaxy's present-day stellar mass is accumulated during each of these phases. In this appendix, we quantify this decomposition for Romeo, our prototypical analogue of the MW, and for the full FIRE sample we work with.

We define the phase masses as the net increase in stellar mass within 20 kpc between the relevant transition times:
\begin{align}
M_{\star,\rm proto} &\equiv M_\star(t_{\rm spin\textnormal{-}up,\star}), \\
M_{\star,\rm thick} &\equiv M_\star(t_{\rm CD}) - M_\star(t_{\rm spin\textnormal{-}up,\star}), \\
M_{\star,\rm thin}  &\equiv M_\star(z=0) - M_\star(t_{\rm CD}),
\end{align}
where $M_\star(t)$ is the stellar mass within 20 kpc of the main progenitor at lookback time $t$. This is an aperture-based measure of stellar mass growth, not a particle-tracked decomposition by stellar birth time. It therefore includes both in situ star formation and accreted stars that enter the main progenitor during a given phase, as well as any net effects from stars moving into or out of the 20 kpc aperture.

This distinction is important. The phase masses defined here should not be interpreted as the present-day masses of kinematically selected protogalaxy, thick disk, or thin disk components. A star added to the galaxy during the thick disk phase need not belong to a present-day thick disk component, and a present-day kinematic thick disk may contain stars formed or accreted over a wider range of times. The numbers reported here should therefore not be compared directly to local estimates of the Milky Way's thick-to-thin disk mass ratio. Instead, they quantify how much stellar mass growth occurred while the galaxy was in each of the three dynamical regimes defined in this paper.

Figure~\ref{fig:romeo_mass_phases} shows the cumulative stellar mass of Romeo within 20 kpc as a function of lookback time, with the three phase intervals shaded. Romeo gains most of its present-day stellar mass during the thick disk phase: $4.71 \times 10^{10}~M_\odot$, corresponding to 64\% of its $z=0$ stellar mass within 20 kpc. The thin disk phase contributes $2.22 \times 10^{10}~M_\odot$, or 30\%, while the protogalaxy phase contributes only $0.44 \times 10^{10}~M_\odot$, or 6\%. This small protogalactic fraction follows from Romeo's early stellar spin-up time, $t_{\rm spin\textnormal{-}up,\star}=11.7$ Gyr, which leaves only a short interval between the onset of star formation and the end of the protogalaxy phase. By contrast, the thick disk phase lasts 5.8 Gyr in Romeo and spans the period over which the stellar mass rises from $\sim 4\times10^9~M_\odot$ to $\sim 5\times10^{10}~M_\odot$.

Figure~\ref{fig:suite_mass_phases} extends this decomposition to all 12 galaxies in the suite. The top panel shows the absolute stellar mass growth in each phase, while the bottom panel shows the same quantities normalized to each galaxy's present-day stellar mass within 20 kpc. The phase breakdown varies widely from galaxy to galaxy. The protogalactic fraction ranges from $\sim 1\%$ in m12m to $\sim 51\%$ in Thelma, the thick disk fraction ranges from $\sim 6\%$ in m12r to $\sim 82\%$ in m12m, and the thin disk fraction ranges from $\sim 15\%$ in m12m to $\sim 65\%$ in m12r. There is no clean monotonic trend with total mass: the most massive system, m12m, and the least massive system, m12r, lie at opposite extremes of the thick-disk versus thin-disk mass-growth breakdown.

The dominant source of this variation is the spread in phase durations. Galaxies with early stellar spin-up and late cooldown spend a large fraction of cosmic history in the thick disk phase, and therefore tend to accumulate most of their stellar mass during that interval. For example, m12m has $t_{\rm spin\textnormal{-}up,\star}=10.94$ Gyr and $t_{\rm CD}=2.56$ Gyr, giving it an extended thick disk phase and the largest thick-disk mass fraction in the suite. Conversely, galaxies in which stellar spin-up and cooldown occur close together accumulate relatively little mass during the thick disk phase. In m12b, for example, $t_{\rm spin\textnormal{-}up,\star}-t_{\rm CD}=0.93$ Gyr, so the stellar mass growth is distributed primarily between the protogalaxy and thin disk phases. Large protogalactic fractions occur in systems with comparatively late stellar spin-up times, such as Thelma and Louise.

The three ELVIS Local Group analog pairs do not show a uniform phase-mass breakdown. Thelma and Louise both have large protogalactic fractions, 51\% and 49\%, respectively, but Romeo and Juliet differ substantially despite sharing the same zoom-in volume, with protogalactic fractions of 6\% and 43\%. Romulus and Remus also differ, though less dramatically. This supports the broader interpretation that the timing of spin-up and cooldown is controlled primarily by each galaxy's own assembly history rather than by the shared large-scale environment of paired systems.

We emphasize, however, that this within-pair diversity does not imply that large-scale environment is irrelevant. \citet{Santistevan2020} showed that FIRE-2 hosts in Local Group-like environments assemble systematically earlier than isolated MW-mass hosts: their main progenitors reach a given fraction of their final stellar mass earlier and reside in dark-matter halos that collapse earlier. Our results are consistent with this broader environmental trend. The point here is more specific: galaxies sharing the same zoom-in volume do not exhibit matched spin-up or cooldown histories. Thus, large-scale environment can shift the typical epoch of assembly, while the detailed timing of the kinematic transitions remains sensitive to each galaxy's individual history.

The robust conclusion from Figure~\ref{fig:suite_mass_phases} is therefore that the stellar mass growth associated with the protogalaxy, thick disk, and thin disk phases varies substantially across the FIRE-2 Milky Way-mass suite. This variation is driven mainly by how long each galaxy spends in each dynamical phase, rather than by present-day total mass alone. The three phases used in this paper are epochs in the dynamical evolution of the main progenitor, not present-day stellar components, and the corresponding mass budget should be interpreted in that sense.

\section{Selection Sensitivity of Simulation Sample}
\label{sec:selection}
\begin{figure*}
    \centering
    \includegraphics[width = \textwidth]{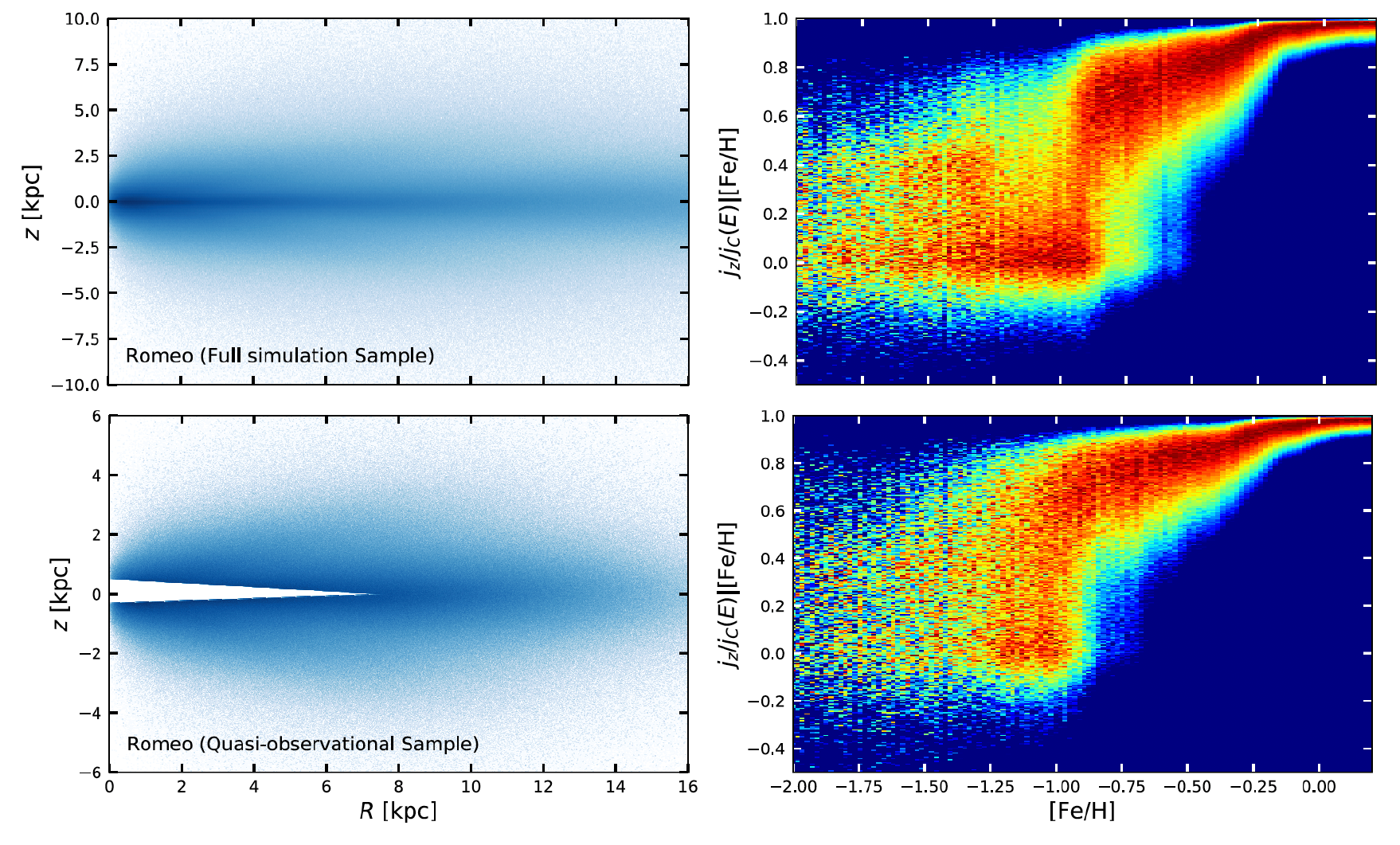}
    \caption{Selection check for the qualitative three-phase comparison between Romeo and the Milky Way. The top row shows all $z=0$ star particles within 20 kpc of the Romeo galaxy center, matching the simple fiducial selection used in Figures~\ref{fig:gaia and romeo three phases} and \ref{fig:jzjc_vphi_feh_stars_gas}. The bottom row shows a more observationally motivated sample chosen to better resemble the spatial character of the \textit{Gaia} RGB sample analyzed by \citet{Chandra2024}. The left panels show the spatial distribution of the selected particles, and the right panels show the corresponding column-normalized distribution of orbital circularity as a function of $[{\rm Fe/H}]$. Although the detailed appearance of the diagram changes with selection, the same qualitative three-phase sequence remains strikingly apparent in both cases.}
    \label{fig:z_R_jzjc_fe_over_h_romeo}
\end{figure*}
In Section~\ref{sec:example}, we compare the three-phase structure seen in the \textit{Gaia} RGB sample of \citet{Chandra2024} to the corresponding structure in Romeo using all $z=0$ star particles within 20 kpc of the simulated galaxy center. This is intentionally not a forward-modeled mock \textit{Gaia} sample. Instead, the purpose of Figure~\ref{fig:gaia and romeo three phases} is to establish a qualitative resemblance between the archaeological circularity--metallicity structure of the Milky Way and that of one representative FIRE-2 Milky Way-mass galaxy. Figure~\ref{fig:z_R_jzjc_fe_over_h_romeo} checks whether that qualitative resemblance depends strongly on this simple stellar selection.

The top row of Figure~\ref{fig:z_R_jzjc_fe_over_h_romeo} shows the same type of circularity--metallicity diagram used in the main text, constructed from all $z=0$ stars within 20 kpc of Romeo. The bottom row instead shows a more observationally motivated sample, chosen to better resemble the spatial character of the \textit{Gaia} RGB sample used by \citet{Chandra2024}. The left panels illustrate the spatial distribution of the selected particles, while the right panels show the corresponding column-normalized distribution of orbital circularity as a function of $[{\rm Fe/H}]$.

As expected, the detailed appearance of the diagram changes with the selection. Different cuts weight the central galaxy, stellar halo, thick disk, and thin disk differently, and therefore alter the relative prominence and sharpness of features at fixed metallicity. These differences are relevant for any future attempt to make a quantitative, survey-matched comparison between FIRE-2 and the \textit{Gaia} data.

For the purposes of this paper, however, the important result is that the qualitative three-phase structure remains visible in both selections. In both the full 20 kpc sample and the more observationally motivated sample, Romeo shows a metal-poor, low-circularity protogalactic component, an intermediate-metallicity population with enhanced prograde rotation and broad dispersion, and a metal-rich population concentrated toward high circularity. Thus, the basic visual resemblance between Romeo and the Milky Way in Figure~\ref{fig:gaia and romeo three phases} is not an artifact of using all stars within 20 kpc.

We therefore use Figure~\ref{fig:z_R_jzjc_fe_over_h_romeo} as a robustness check on the qualitative comparison made in Section~\ref{sec:example}. A fully realistic mock observation would require modeling the \textit{Gaia} selection function, the RGB selection, survey geometry, distance uncertainties, and observational errors. That is beyond the scope of this work. Our conclusions instead rely on the fact that the same broad three-phase sequence still appears under reasonable stellar selections.

\section{Maximum Circular Velocities as a Function of Lookback Time}
\label{sec:v_max}

\begin{figure}
    \centering
    \includegraphics[width=\columnwidth]{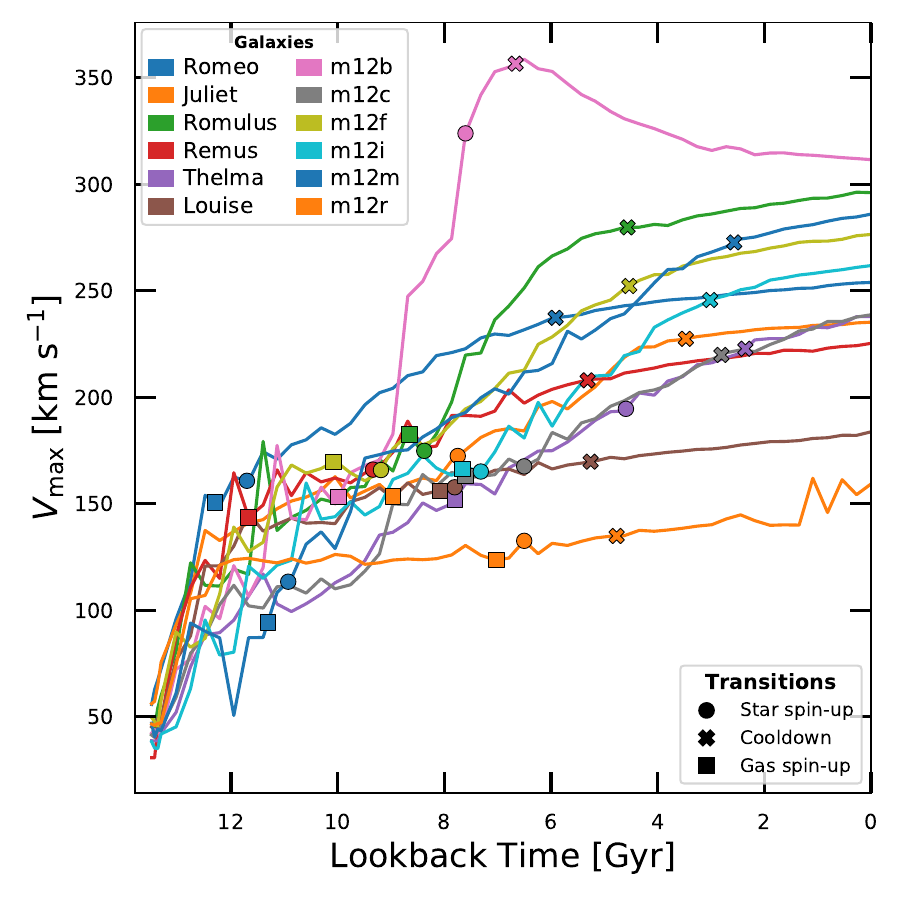}
    \caption{Maximum circular velocity, $V_{\rm max}$, as a function of lookback time for each galaxy in our simulation suite. Symbols mark the gas spin-up time (squares), stellar spin-up time (circles), and cooldown time ($\times$) defined in Section~\ref{sec:sample-results}. The transitions occur over broad and overlapping ranges of $V_{\rm max}$, indicating that none of the three kinematic transitions is associated with a single characteristic maximum circular velocity.}
    \label{fig:v_max}
\end{figure}

\autoref{fig:v_max} shows the maximum circular velocity, $V_{\rm max}$, as a function of lookback time for each galaxy in our suite, with markers denoting the gas spin-up (squares), stellar spin-up (circles), and cooldown ($\times$) times defined in Section~\ref{sec:sample-results}. $V_{\rm max}$ provides a complementary single-number summary of the characteristic circular velocity of each main progenitor, and is more directly tied to disk orbital speeds than the virial and stellar masses shown in \autoref{fig:characteristic masses and gas fraction}. As with the mass tracks in \autoref{fig:characteristic masses and gas fraction}, $V_{\rm max}$ rises rapidly at early times and grows more slowly toward $z = 0$, consistent with the canonical halo growth picture \citep{Wechsler02}. The clearest outlier is m12b, which experiences a sharp jump of $\sim 200~{\rm km\,s^{-1}}$ near a lookback time of $\sim 7$~Gyr associated with a major merger; m12r, by contrast, remains among the lowest-$V_{\rm max}$ systems throughout its history.

We find that gas and stellar spin-ups typically occur at $V_{\rm max} \approx 120$--$180~{\rm km\,s^{-1}}$, while cooldown occurs later and at substantially higher values, $V_{\rm max} \approx 200$--$280~{\rm km\,s^{-1}}$ for most systems, with m12b reaching $\sim 355~{\rm km\,s^{-1}}$. The three sets of markers occupy roughly distinct but overlapping regions of the diagram, with substantial galaxy-to-galaxy scatter at each transition. As with the masses in \autoref{fig:characteristic masses and gas fraction}, no sharp characteristic value of $V_{\rm max}$ coincides with any of the three transitions: the gas and stellar spin-up markers span a broad range, and overlap with the cooldown values of the lowest-$V_{\rm max}$ systems.

\section{Example Velocity Map of Kinematic Phases}
\label{sec:velmap}

\begin{figure*}
    \centering
    \includegraphics[height=\dimexpr\textheight-2.5cm\relax]{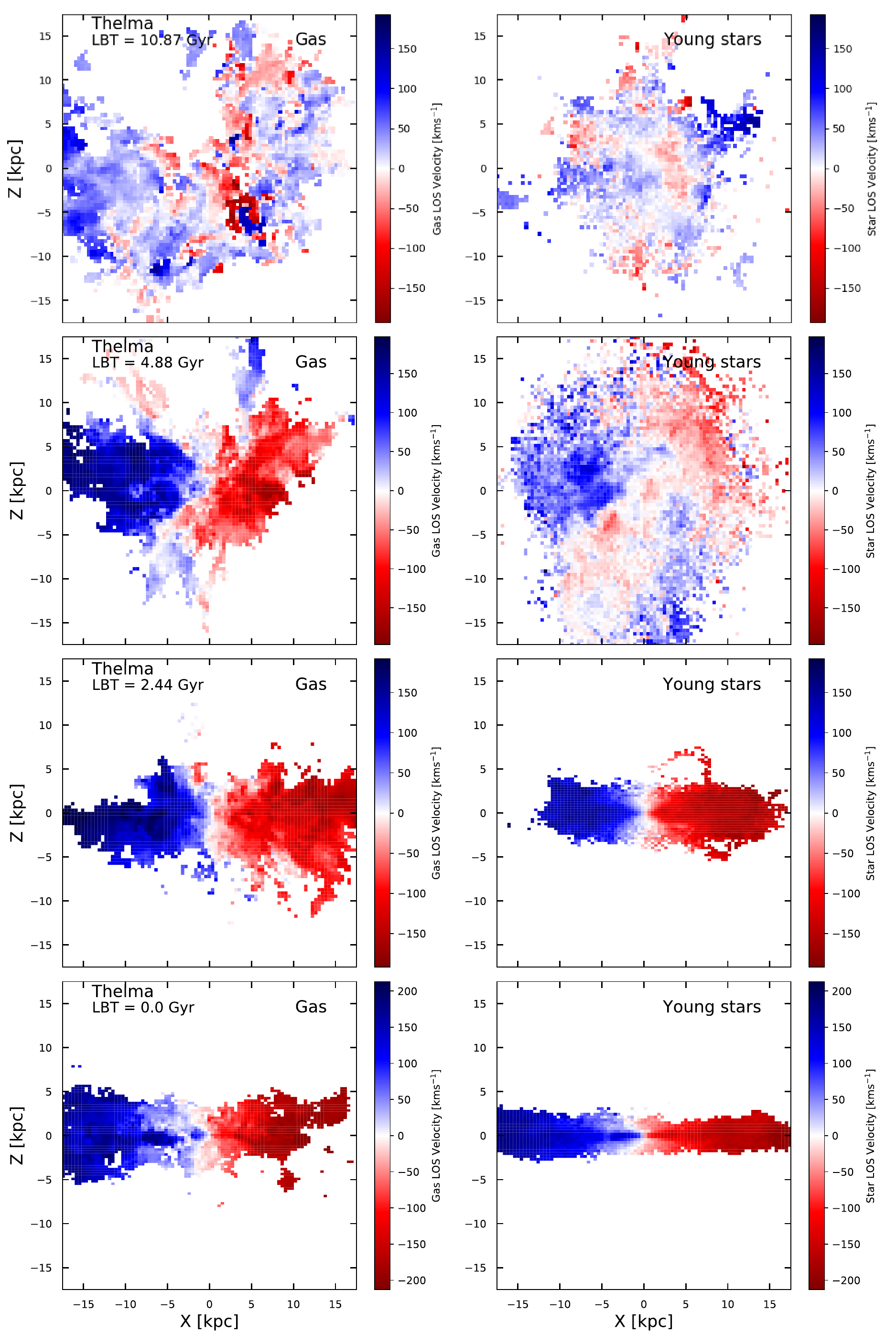}
    \caption{Distinguishing the separate components of a galaxy disk through edge-on velocity field maps. The gas disk forms first, followed by a stellar disk. The top panel is during the protogalaxy era (before the formation of any disk), the second panel is just before stellar spin-up and the third panel is just before cooldown. The bottom panel shows the velocity field at $z = 0$. We remove gas with temperature above $10^4$ K in all panels. Gas coordinates and velocities are rotated with respect to the net angular momentum of this cool gas, while star coordinates and velocities are rotated with respect to the net angular momentum of young stars ($\leq 250$ Myr in age).}
    \label{fig:velocity field maps}
\end{figure*}

Figure~\ref{fig:velocity field maps} provides a visual illustration of the kinematic phases discussed throughout the paper, using Thelma as an example system. Shown are edge-on projected velocity field maps for cool gas and young stars at four representative times. LBT refers to lookback time. Images on the left show cool gas, selected with $T < 10^4$ K, while images on the right show young stars, selected to have ages $<250$ Myr at each snapshot. In each panel, the system is rotated into an edge-on orientation using the angular momentum direction of the component being shown: gas maps are oriented with respect to the angular momentum of the cool gas, while young-star maps are oriented with respect to the angular momentum of the young stars. The color scale indicates the line-of-sight velocity field.

The rows are chosen to correspond to the sequence of transition times defined in Section~\ref{sec:sample-results}. The top row shows an early snapshot before gas spin-up, during the protogalaxy phase. At this time, neither the cool gas nor the young stars display a clean, ordered velocity gradient. The gas is morphologically irregular and kinematically disordered, while the young-star velocity field is similarly incoherent. This visual impression is consistent with the low median circularities seen before spin-up in Figures~\ref{fig:romeo median and 1sigma} and \ref{fig:fire suite jzjc medians}.

The second row shows a snapshot after gas spin-up but before stellar spin-up. This is the phase in which the gas has begun to acquire coherent rotation, but the newly forming stars have not yet settled into the same ordered kinematic configuration. The cool gas map shows a clearer velocity gradient than in the first row, while the young-star map remains comparatively disordered. This illustrates the time delay emphasized in Figures~\ref{fig:fire suite jzjc medians} and \ref{fig:t_gas t_star t_cool}: gas spin-up typically precedes stellar spin-up.

The third row shows a snapshot after stellar spin-up but before cooldown, corresponding to the thick disk phase. By this stage, both gas and young stars show coherent rotation, but the young-star component remains vertically extended and kinematically hotter than the final thin disk. This is the regime in which the median circularity has risen above the protogalactic value but the distribution remains broad, so newly formed stars occupy disk-like orbits without yet being confined to a dynamically cold thin disk.

The bottom row shows the $z=0$ system, after cooldown. Both the cool gas and the young stars now display ordered rotation in a geometrically thin configuration. The velocity gradients are visually coherent and the young-star component is much more disk-like than in the pre-cooldown rows. This final row therefore provides a spatial and kinematic counterpart to the high-circularity, low-dispersion thin disk phase identified in the time-domain circularity analysis.

Overall, Figure~\ref{fig:velocity field maps} shows the same sequence inferred quantitatively from the circularity evolution: an initially disordered protogalaxy, followed by gas spin-up, then stellar spin-up into a thick disk, and finally cooldown into a thin disk. The figure also makes visually clear why gas and stars are treated as separate tracers in our transition-time definitions. The gas can develop an ordered velocity field before the young stars do, while the final thin disk phase requires both components to become coherently rotating and geometrically settled.

\section{Gas and Young Star Morphologies at Spin-up}
\label{sec: mergers}
\begin{figure*}
    \centering
    \includegraphics[width = \textwidth]{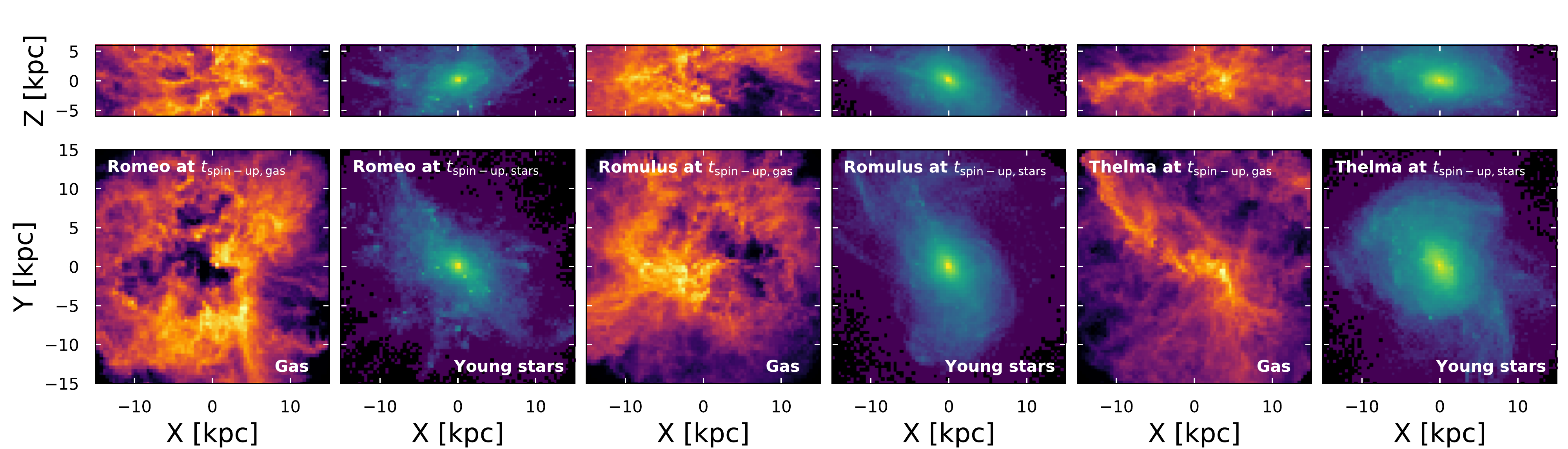}
    \caption{Projected gas and young-star (less than $250$ Myr old) morphological views for Romeo, Romulus, and Thelma at their respective spin-up times. For each galaxy, the gas panel shows the morphology at $t_{\rm spin-up,gas}$, while the young-star panel shows the morphology at $t_{\rm spin-up,\star}$. Top panels show edge-on projections and bottom panels show face-on projections.}
    \label{fig:mergers}
\end{figure*}
Figure \ref{fig:mergers} shows projected gas and young-star ($t_\star < 250$ Myr) density distributions for Romeo, Romulus, and Thelma at their respective gas and stellar spin-up times, in both face-on and edge-on projections. Across the three systems, the gas is clumpy and filamentary at spin-up, consistent with the chaotic accretion environment expected at early cosmic times. We discuss the possible contribution of such clumpy gas and its associated feedback to early baryonic disequilibrium in Section \ref{subsec:sloshing}.\par

Romulus displays the most disturbed morphology: a coherent gas stream extends across the inner $\sim 30$ kpc, and the young-star distribution is elongated and asymmetric, consistent with the discrete mass accretion event identified for Romulus in Section \ref{sec:empirical-trends}. Romeo, by contrast, shows turbulent and clumpy gas without an obvious large-scale stream, along with a smooth, centrally concentrated young-star core. Thelma displays a clear accreting gas filament, but otherwise shows a smooth stellar morphology similar to Romeo.\par

Importantly, none of the three systems exhibits an obvious comparable-mass stellar merger at the time of spin-up. Combined with the absence of systematic mass jumps coincident with the kinematic transitions in the full sample (Section 4.1; Figure 7), these morphologies support the interpretation that disk formation proceeds within a broadly chaotic 
accretion environment rather than being triggered by discrete major mergers.

\section{Efficiency of Gas Consumption Across Different Temperature Thresholds}
\label{sec:gce temps}

\begin{figure*}
    \centering
    \includegraphics[width = \textwidth]{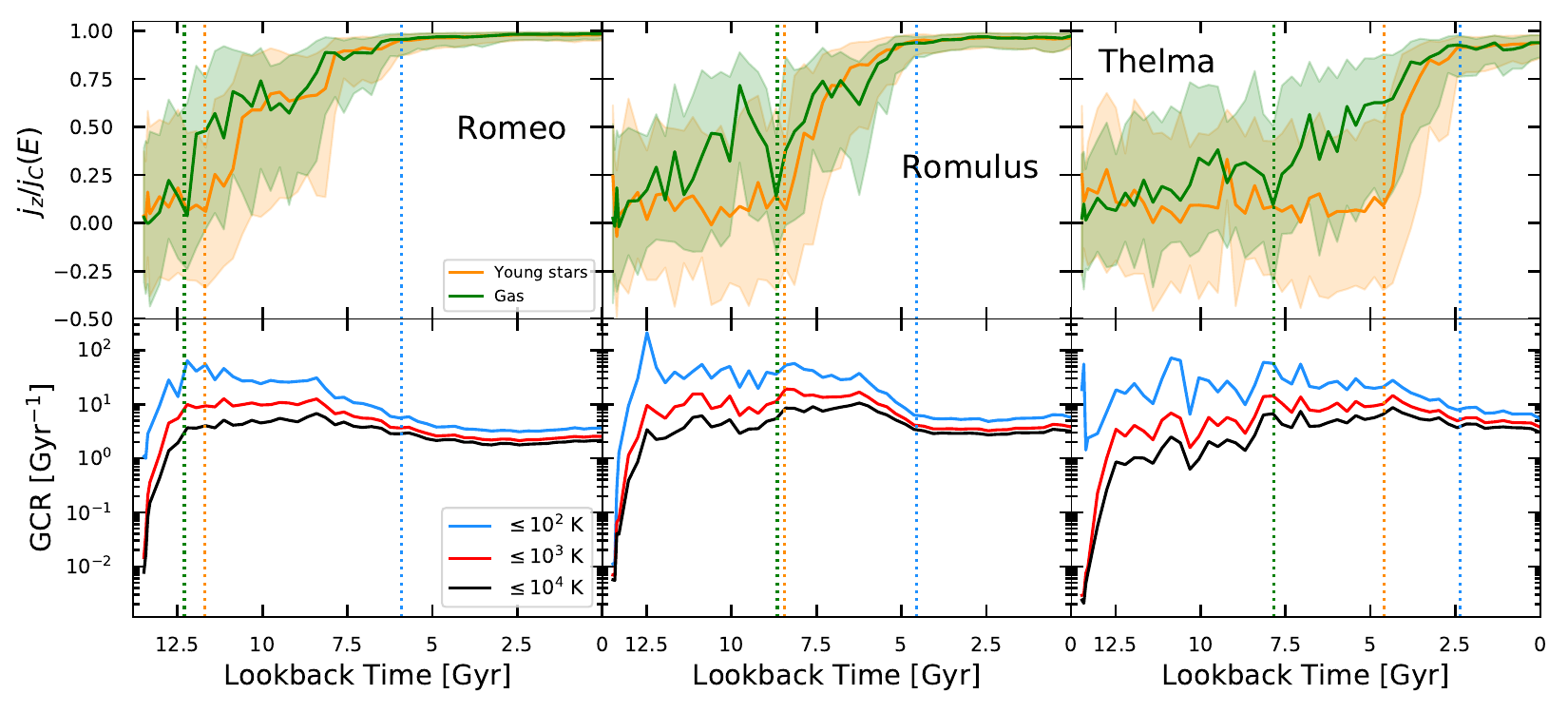}
    \caption{Gas consumption rate as a function of temperature threshold for Romeo (left), Romulus (middle), and Thelma (right). \textbf{Top row:} Median orbital circularity $j_z / j_c(E)$ of gas (green) and young stars (orange, $< 250$~Myr old at each snapshot) as a function of lookback time. Shaded bands show the 25th-to-75th-percentile range. \textbf{Bottom row:} Gas consumption rate $\mathrm{GCR} \equiv \dot{M}_\star / M_\mathrm{gas}(T \leq T_\mathrm{thresh})$, computed with a 250~Myr-averaged SFR, for three temperature thresholds: $T \leq 10^4$~K (black; the cool-gas definition used in Eq.~\ref{eq:gcr} and Figure~\ref{fig:starformation}), $T \leq 10^3$~K (red), and $T \leq 10^2$~K (blue). Vertical dotted lines in every panel mark the gas spin-up time (green), the stellar spin-up time (orange), and the cooldown time (blue) for each galaxy. At fixed lookback time, the GCR is systematically higher for colder thresholds; the colder the threshold, the earlier the GCR peaks and the more pronounced its decline following cooldown.}
    \label{fig:plot_jzjc_gcr_temps}
\end{figure*}

In Section~\ref{subsec:star-formation}, we defined the gas consumption rate (GCR; Eq.~\ref{eq:gcr}) as the 250 Myr-averaged star formation rate divided by the mass of cool gas, $T \leq 10^4~{\rm K}$, within 20 kpc of the galaxy center. This definition measures how rapidly the broad cool-gas reservoir is converted into stars. However, because star formation occurs only in the densest and coldest subset of this material, normalizing by all $T \leq 10^4~{\rm K}$ gas may obscure how the more directly star-forming gas reservoir evolves across the three kinematic phases. To test the sensitivity of our results to this choice of denominator, we repeat the calculation using two colder temperature thresholds:
\begin{equation}
{\rm GCR}_{T_{\rm max}}(t)
=
\frac{{\rm SFR}_{250~{\rm Myr}}(t)}
{M_{\rm gas}(T \leq T_{\rm max}, t)},
\end{equation}
with $T_{\rm max}=10^4,10^3,$ and $10^2~{\rm K}$.

Figure~\ref{fig:plot_jzjc_gcr_temps} shows this comparison for Romeo, Romulus, and Thelma. The top row repeats the evolution of the median orbital circularity, $j_z/j_c(E)$, for gas and young stars, with vertical dotted lines marking the gas spin-up time, stellar spin-up time, and cooldown time. The bottom row shows ${\rm GCR}$ as a function of three temperature thresholds. The $T \leq 10^4~{\rm K}$ curve is identical to the cool-gas GCR used in Section~\ref{subsec:star-formation}, while the $T \leq 10^3~{\rm K}$ and $T \leq 10^2~{\rm K}$ curves progressively isolate colder gas.

Two trends are apparent across all three galaxies. First, at fixed lookback time, the inferred GCR increases as the temperature threshold is lowered. This is expected: colder temperature cuts select a smaller gas reservoir that is more closely associated with the material participating in star formation, so the same SFR corresponds to a larger consumption rate. The $T \leq 10^2~{\rm K}$ curves are typically about an order of magnitude higher than the $T \leq 10^4~{\rm K}$ curves. Second, the colder-gas GCRs tend to peak earlier and decline more strongly after cooldown. This behavior is clearest in Thelma, where the $T \leq 10^2~{\rm K}$ curve reaches values of several tens of ${\rm Gyr}^{-1}$ before cooldown and then falls to a few ${\rm Gyr}^{-1}$ by $z=0$. The corresponding $T \leq 10^4~{\rm K}$ curve peaks later, at lower amplitude, and undergoes a more modest post-cooldown decline. Romeo and Romulus show the same ordering with temperature threshold, although the peak times of the three curves are less widely separated.

These trends suggest that the post-cooldown reduction in gas consumption rate is strongest when measured relative to the coldest gas reservoir. Equivalently, after cooldown the galaxy maintains a larger reservoir of cold gas per unit star formation rate than it did during the burstier pre-cooldown phases. This is consistent with the physical picture discussed in Section~\ref{subsec:star-formation}: once the galaxy settles into a rotationally supported thin disk, cold gas clouds are embedded in a more ordered shear flow and may be stretched or stabilized before collapsing into self-gravitating star-forming regions. In addition, previous work with FIRE-2 has shown that the thick-to-thin disk transition coincides with a transition from bursty to steady star formation \citep{Yu2021, Stern_2021, Gurvich23}. In this steady late-time regime, stellar feedback regulates the global star formation efficiency per free-fall time to values well below the local efficiency assumed for self-gravitating star-forming gas \citep{CAFG2013, Hopkins17, Orr18}. Both effects would act most directly on the dense, cold gas, naturally producing a larger drop in the ${10^2~{\rm K}}$ ${\rm GCR}$ than in the ${10^4~{\rm K}}$ ${\rm GCR}$.

We emphasize, however, that these temperature cuts should be interpreted as approximate diagnostics rather than clean phase selections. In FIRE-2, star-forming gas is required to be locally self-gravitating, dense, Jeans unstable, and molecular \citep{Krumholz2011}, but a cut on temperature alone is not identical to a selection on molecular, self-gravitating, or gravitationally bound gas. The absolute normalization of the coldest-gas GCR, especially at early times when the cold gas mass can be small and bursty, should therefore be interpreted with caution. The robust result is the relative ordering: colder gas thresholds yield higher GCRs, earlier peaks, and a more pronounced decline after cooldown.

\section{More on the Concentration of the Potential}
\label{sec: central concentration appendix}

\begin{figure*}
\centering
\includegraphics[width = \textwidth]{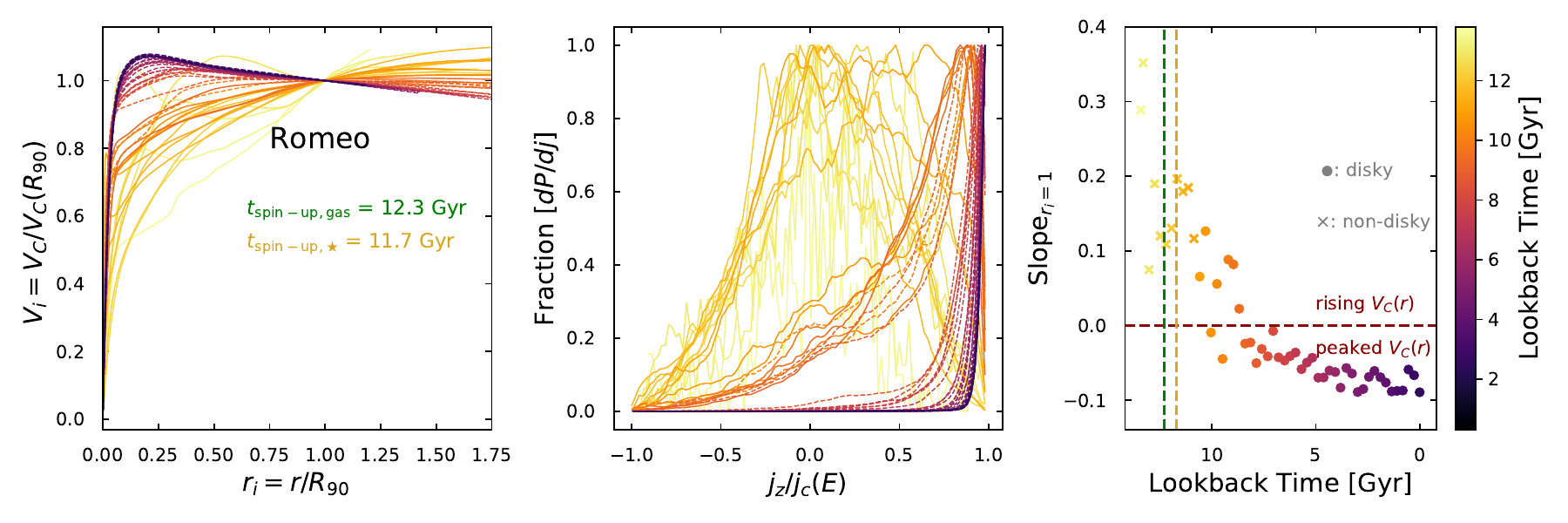}
\caption{Time evolution of the rotation-curve shape, young-star circularity, and a slope-based concentration metric for Romeo. \textit{Left:} Normalized circular velocity profile $V_C/V_C(R_{90})$ versus $r/R_{90}$ at each cosmic snapshot, color-coded by lookback time. Solid lines mark snapshots with a positive slope at $r_i \equiv r/R_{90} = 1$ (a still-rising rotation curve at $R_{90}$, indicating an extended or weakly concentrated potential), while dashed lines mark snapshots with a negative slope (a rotation curve that has already peaked interior to $R_{90}$, indicating a more centrally concentrated potential). \textit{Middle:} Distributions of orbital circularity $j_z/j_c(E)$ for young stars (younger than 250 Myr at each snapshot), with the same color coding and line-style convention as the left panel. The distributions evolve from broad and nearly isotropic at early times to sharply peaked near $j_z/j_c = 1$ at late times, tracing the buildup of coherent stellar rotation. \textit{Right:} Slope of the normalized rotation curve at $r_i = 1$ as a function of lookback time. Dots indicate snapshots in which at least 61\% of the young-star circularity distribution falls within $0.2 \leq j_z/j_c \leq 1$, our operational definition of a disky distribution; crosses indicate non-disky snapshots. The horizontal dashed red line marks zero slope, separating rising from peaked rotation-curve configurations. The vertical dashed green and goldenrod lines mark the gas and stellar spin-up times for Romeo, respectively. The slope reaches its maximum positive value near spin-up and crosses zero only several Gyr later, after a coherently rotating stellar component is already established.}
\label{fig:plot_jzjc_circ_vel}
\end{figure*}

In Section~\ref{sec: concentration}, we used the ratio $R_{\rm peak}/20{\rm kpc}$ as a measure of the central concentration of the gravitational potential. Here, we present two complementary checks based directly on the rotation curve. First, we examine the local logarithmic slope of the normalized rotation curve, $V_C/V_C(R_{90})$, at the characteristic radius $r_i \equiv r/R_{90}=1$. A negative slope at $r_i=1$ corresponds to a rotation curve that peaks interior to $R_{90}$, indicating a centrally concentrated potential. A positive slope corresponds to a rotation curve that is still rising at $R_{90}$, indicating a more extended or weakly concentrated potential. Second, we compare our FIRE-2 galaxies to the disk-formation criterion recently proposed by \citet{semenov2026}, who argued that gas disks form once the potential becomes steep enough that the circular velocity at the baryonic half-mass scale exceeds the characteristic turbulent velocity of the accreting gas. In the cold-mode regime, where $\sigma \sim V_{\rm vir}$, this condition corresponds approximately to
\begin{equation}
\frac{V_C(R_{50,b})}{V_C( R_{\rm vir})} \gtrsim 1.
\end{equation}
We measure this quantity by defining $R_{50,b}$ as the radius enclosing half of the baryonic mass (gas plus stars) within 20~kpc.

Figure~\ref{fig:plot_jzjc_circ_vel} illustrates the slope-based diagnostic for Romeo, our example MW-mass galaxy from the main text. The three panels show, from left to right, the normalized rotation curves $V_C/V_C(R_{90})$ versus $r/R_{90}$ at each snapshot, the corresponding young-star circularity distributions, and the slope of the normalized rotation curve at $r_i=1$ as a function of lookback time. Solid lines in the left and middle panels indicate snapshots with a positive slope at $r_i=1$, while dashed lines indicate snapshots with a negative slope. In the right panel, dots indicate snapshots satisfying our operational disky criterion, and crosses indicate non-disky snapshots. The vertical dashed green and gold lines mark the gas and stellar spin-up times for Romeo at 12.3 and 11.7 Gyr lookback time, respectively.

The key feature of Figure~\ref{fig:plot_jzjc_circ_vel} is the time-ordering. In Romeo, the slope at $r_i=1$ reaches its maximum positive value near the spin-up time, when the rotation curve is still rising throughout the galaxy. The slope then decreases steadily as the rotation curve transitions from rising to peaked, crossing zero only several Gyr after spin-up has already occurred. Meanwhile, the young-star circularity distributions begin to transition from broad and nearly isotropic to disky while the rotation curve is still rising. Thus, in Romeo, the initial onset of coherent stellar rotation does not occur when the potential has become centrally concentrated. Rather, by this metric, the disk begins to form when the potential is still quite unconcentrated, and the strongest central concentration develops afterward during the thick-disk phase.

\begin{figure*}
\centering
\includegraphics[width = \textwidth]{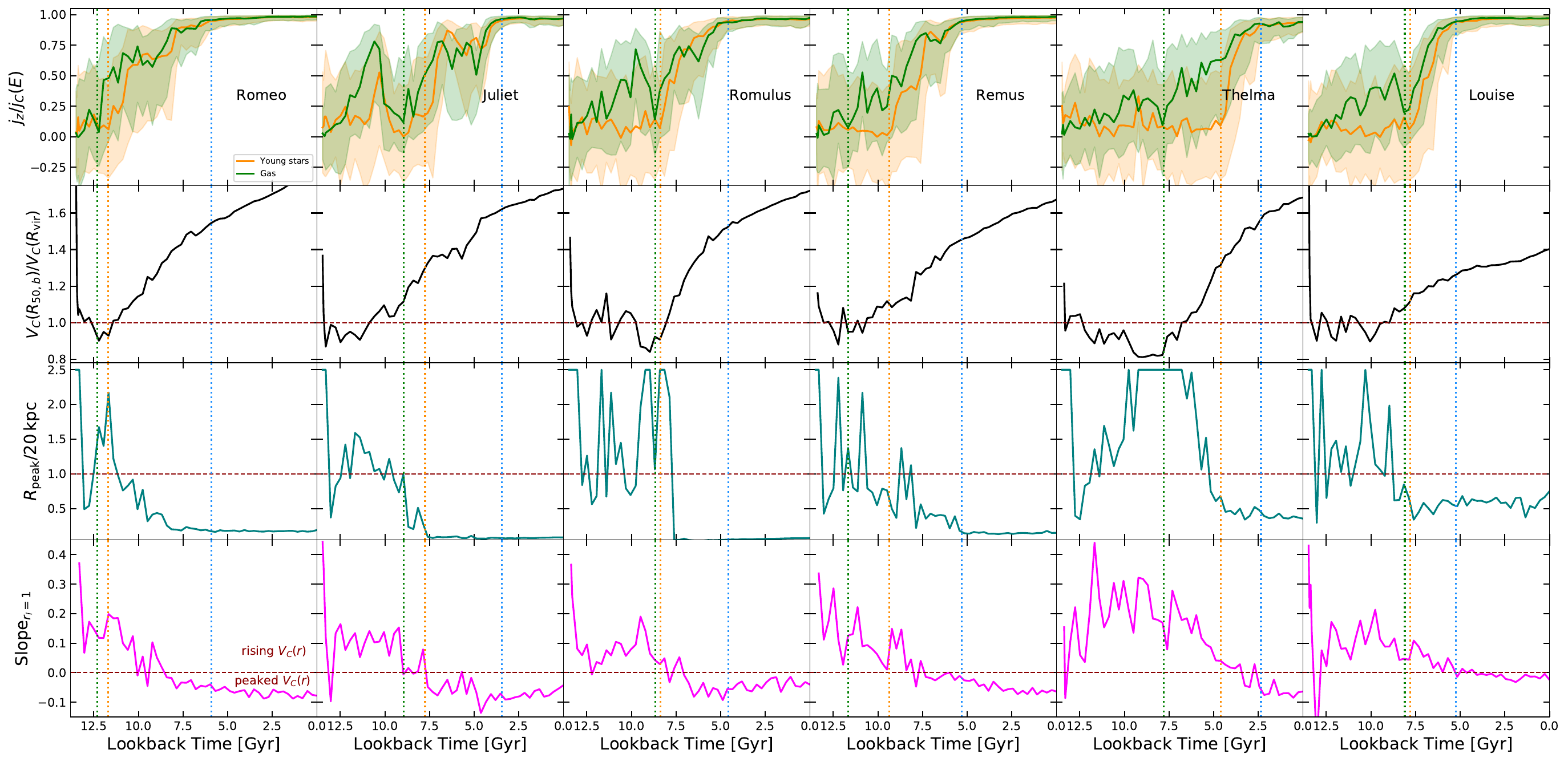}
\caption{ Comparison of central-concentration parameters with the kinematic transition times for the six ELVIS galaxies. Each column corresponds to one galaxy. \textit{First row:} median orbital circularity $j_z/j_C(E)$ of young stars ($<250$~Myr; orange) and gas (green), with shaded $16$th--$84$th percentile bands. \textit{Second row:} the parameter $V_C(R_{50,b})/V_C(R_ {\rm vir})$ introduced in \citet{semenov2026}, where $R_{50,b}$ is the baryonic half-mass radius measured within 20~kpc. The dashed line marks unity. \textit{Third row:} $R_{\rm peak}/20{\rm kpc}$, with values above unity indicating a rotation curve that is still rising across the inner 20~kpc. \textit{Fourth row:} the local slope of the normalized rotation curve at $r_i\equiv r/R_{90}=1$, with positive values indicating a rising rotation curve and negative values indicating a peaked rotation curve. Vertical dotted lines mark gas spin-up (green), stellar spin-up (orange), and cooldown (blue). The \citet{semenov2026} parameter is typically near unity (or even at a local minimum below unity) around gas spin-up and rises afterward, while $R_{\rm peak}/20 {\rm kpc}$ and the rotation-curve slope indicate that the potential is unconcentrated near spin-up and steepens during the subsequent thick disk.}
\label{fig:semenov_criterion_elvis}
\end{figure*}

Figure~\ref{fig:semenov_criterion_elvis} extends this comparison to the six ELVIS galaxies. From top to bottom, the rows show the circularity evolution of gas and young stars, the \citet{semenov2026} steepness parameter $V_C(R_{50,b})/V_{\rm vir}$, the $R_{\rm peak}/20 {\rm kpc}$ concentration metric used in Figure~\ref{fig:plot_jzjc_circ_vel}, and the local slope of the normalized rotation curve at $r_i=1$. The vertical dotted lines mark the gas spin-up, stellar spin-up, and cooldown times for each galaxy.

The three concentration diagnostics give a broadly consistent picture. In most of the ELVIS galaxies, $V_C(R_{50,b})/V_C(R_{\rm vir})$ is near unity around the gas spin-up time, often close to a local minimum, and then rises substantially during the thick-disk phase. This behavior is mirrored by $R_{\rm peak}/20 {\rm kpc}$, which tends to be large near spin-up, indicating that the rotation curve is still rising across the inner 20~kpc, and then decreases as the rotation curve becomes more centrally peaked. The slope-based diagnostic shows the same ordering: the slope at $r_i=1$ is generally positive around spin-up and declines toward zero or negative values only later. Thus, across the ELVIS subset, the potential is not particularly concentrated when coherent gas rotation first appears. Instead, the strongest steepening of the central potential occurs after gas spin-up, during the thick disk phase.

In our FIRE-2 sample, the ratio often only rises after the spin-up of young stars. Moreover, its post-spin-up increase includes the response of the baryonic distribution to disk growth itself. In this respect, it is not a purely external predictor of disk formation. Thus, a rising $V_C(R_{50,b})/V_C(R_{\rm vir})$ after spin-up may reflect baryonic contraction associated with the formation and growth of the disk, rather than an independent cause of the initial spin-up event.

We therefore interpret the increase in central concentration as most closely tied to the maturation of the disk rather than to the initial onset of coherent gas rotation. The initial spin-up transition occurs when the potential is near the boundary between an extended, rising rotation curve and a more centrally peaked one. Subsequently, the growth of the disk then steepens the potential further.

\section{Baryonic Sloshing Across the Full Simulation Suite}
\label{sec:offset_full_suite}

\begin{figure*}
    \centering
    \includegraphics[height=\dimexpr\textheight-3.5cm\relax]{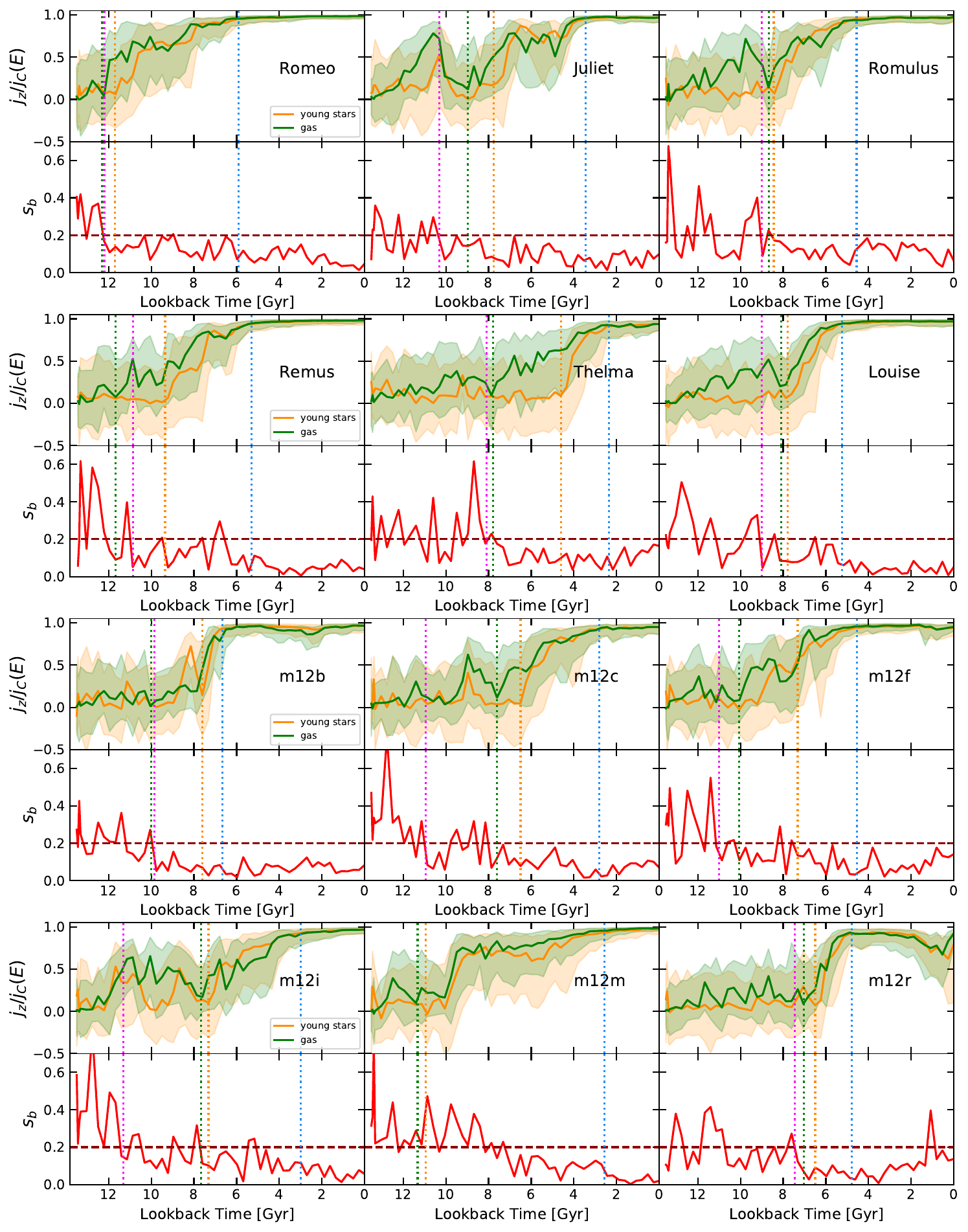}
    \caption{ Baryonic sloshing parameter $s_{\rm b}$ across the full simulation suite. For each of the 12 galaxies in our sample, the upper sub-panel shows the median orbital circularity $j_z/j_C(E)$ of young stars (orange; $<250$ Myr old at each snapshot) and gas (green) as a function of lookback time, with shaded bands indicating the central 90th-percentile range. The lower sub-panel shows the time evolution of the baryonic sloshing parameter $s_{\rm b}$ (red; Eq.~\ref{baryon sloshing}). The horizontal dashed red line marks the reference value $s_{\rm b}=0.2$. Vertical dotted lines indicate the gas spin-up time (green), stellar spin-up time (orange), cooldown time (blue), and sloshing time $t_{\rm slosh}$ (magenta), defined as the lookback time at which $s_{\rm b}$ first crosses below 0.2. In most systems, $s_{\rm b}$ declines to small values near the gas spin-up time, while short-lived late-time excursions, such as in m12r, indicate subsequent dynamical perturbations.}
    \label{fig:sb_full_suite}
\end{figure*}

In Section~\ref{subsec:sloshing} we introduced the baryonic sloshing parameter $s_{\rm b}$ (Eq.~\ref{baryon sloshing}) as a measure of the offset between the bulk velocity of the baryons within 20 kpc and the total center-of-mass velocity of the system, normalized by the circular velocity at $R_{90}$. Figure~\ref{fig:jzjc tdyn baryon} showed the time evolution of $s_{\rm b}$ for Romeo, Romulus, and Thelma, demonstrating that $s_{\rm b}$ tends to fall below $\sim 0.2$ around the time of gas spin-up in these example systems. Figure~\ref{fig:t_slosh_t_s,gas} summarized this behavior across the suite by comparing the lookback time at which $s_{\rm b}$ first crosses below 0.2 to the gas spin-up time. For completeness, this appendix shows the full time evolution of $s_{\rm b}$ alongside the circularity evolution for all 12 galaxies in our sample.

Figure~\ref{fig:sb_full_suite} presents, for each galaxy, the median circularity $j_z/j_C(E)$ of young stars (orange) and gas (green) in the upper sub-panel, and the time evolution of $s_{\rm b}$ (red) in the lower sub-panel. The horizontal dashed red line in each lower sub-panel marks the reference value $s_{\rm b}=0.2$. The vertical dotted green, orange, blue, and magenta lines mark the gas spin-up, stellar spin-up, cooldown, and sloshing times, respectively, as defined in Section~\ref{sec:sample-results} and Section~\ref{sec:causation}.

Across the suite, $s_{\rm b}$ exhibits the same qualitative behavior seen in the three example systems shown in Figure~\ref{fig:jzjc tdyn baryon}: large-amplitude and chaotic, erratic jumps at early times, followed by a decline to smaller values as the gas disk begins to form. In most galaxies, the transition from frequent excursions above 0.2 to a quieter sub-threshold state occurs at or shortly before the gas spin-up time. After spin-up, typical values of $s_{\rm b}$ are well below the 0.2 reference value, often in the range $\sim 0.05$--$0.15$, indicating that the baryons and the total mass distribution move more coherently once a disk has formed.

Remus and m12m provide exceptions to a simple one-to-one correspondence between the first crossing of $s_{\rm b}=0.2$ and gas spin-up. Although the median gas circularity in Remus rises at the adopted gas spin-up time, $s_{\rm b}$ continues to fluctuate near and above the reference threshold for several Gyr afterward (visible as the offset between the green and magenta dotted lines in the Remus panel). This behavior illustrates why we treat $s_{\rm b}=0.2$ as an empirical marker rather than a sharp physical boundary. Short-timescale fluctuations in $s_{\rm b}$ can be driven by mergers, bursty outflows, or incoming gas streams, so the broader decline in sloshing amplitude across the spin-up era is more robust than the exact time of first threshold crossing.

A second feature worth noting is that short-lived spikes in $s_{\rm b}$ can occur even after cooldown. This is most apparent in m12r, which experiences a major merger about 1 Gyr prior to $z=0$ (see Section~\ref{sec:sample-results}). The corresponding late-time rise in $s_{\rm b}$ is consistent with a discrete dynamical perturbation of an already-formed disk. Thus, $s_{\rm b}$ should be interpreted as a measure of the stability of the baryonic center of mass, not as a binary indicator of whether a disk is present.

Taken together, Figure~\ref{fig:sb_full_suite} shows that the decline of baryonic sloshing is closely and consistently associated with the onset of coherent gas rotation across the suite, with m12m as the only galaxy in which the timing of the first threshold crossing departs significantly from the gas spin-up time (to be clear, Remus also has $t_{\rm spin-up, gas}> t_{\rm slosh}$, but only by 1 Gyr). This supports the interpretation developed in Section~\ref{subsec:sloshing} that the formation of a stable baryonic center of mass is the most consistent prerequisite for the spin-up transition identified in our suite.




%

\bsp	
\label{lastpage}
\end{document}